\documentclass[aps,prb,showpacs,revsymb,amsfonts,amssymb,twocolumn,
groupedaddress,footinbib,floatfix]{revtex4-1}

\usepackage{graphicx}
\usepackage{amsfonts}
\usepackage{amssymb}
\usepackage{amsmath}
\usepackage{verbatim}
\usepackage{yfonts}

\usepackage[pdftex,bookmarks,colorlinks,breaklinks]{hyperref}  
\hypersetup{linkcolor=blue,citecolor=red,filecolor=green,urlcolor=green} 

\newcommand{\mbf}[1]{{\bf #1}}

\newcommand{\myu}{u}

\begin{document}

\title{Vortex dynamics in type II superconductors under strong pinning
conditions}

\author{A.U.\ Thomann}
\author{V.B.\ Geshkenbein}
\author{G.\ Blatter}
\affiliation{Institute for Theoretical Physics, ETH Zurich, 
8093 Z\"urich, Switzerland} 

\date{\today}

%
\begin{abstract}
We study effects of pinning on the dynamics of a vortex lattice in a type II
superconductor in the strong-pinning situation and determine the
force--velocity (or current--voltage) characteristic combining analytical and
numerical methods. Our analysis deals with a small density $n_p$ of defects
that act with a large force $f_p$ on the vortices, thereby inducing bistable
configurations that are a characteristic feature of strong pinning theory.  We
determine the velocity-dependent average pinning-force density $\langle
F_p(v)\rangle$ and find that it changes on the velocity scale $v_p \sim
f_p/\eta a_0^3$, where $\eta$ is the viscosity of vortex motion and
$a_0$ the distance between vortices. In the small pin-density limit, this
velocity is much larger than the typical flow velocity $v_c \sim F_c/\eta$ of
the free vortex system at drives near the critical force-density $F_c =
\langle F_p(v=0)\rangle \propto n_p f_p$. As a result, we find a generic
excess-force characteristic, a nearly linear force--velocity characteristic
shifted by the critical force-density $F_c$; the linear flux-flow regime is
approached only at large drives. Our analysis provides a derivation of
Coulomb's law of dry friction for the case of strong vortex pinning.
\end{abstract}
%
\maketitle
%

\section{Introduction \label{sec:introduction}}

Superconductors carry electric current without dissipation \cite{onnes_11} and
expell magnetic fields from their body, known as the Meissner-Ochsenfeld
effect \cite{meissner_33}.  In a type II superconductor, magnetic fields $H$
in the range between the lower ($H_{c1}$) and upper ($H_{c2}$) critical fields
penetrate the material in the form of quantized flux lines ($\Phi_0 = hc/2e$)
or vortices, resulting in the mixed or Shubnikov \cite{shubnikov:37} phase.
The repulsive interaction mediated by the vortex currents leads to the
formation of an Abrikosov vortex lattice  \cite{abrikosov_57} with an average
induction $B$ inside the sample.  External currents $j$ drive the vortices
through the Lorentz force density $F_{\rm \scriptscriptstyle L} = j B/c $,
giving rise to vortex motion and dissipation. The vortex velocity $v$ is
determined by the force balance equation $\eta v = F_{\rm\scriptscriptstyle
L}$ with the Bardeen-Stephen viscous coefficient \cite{bardeen_65} $\eta \sim
B H_{c2}/ \rho_n c^2$ and $\rho_n$ the normal state resistivity. The resulting
electric field $E= B v/c$ deprives the superconductor from its defining
property, to carry electric current without dissipation, with the emerging
linear response characterized by the flux-flow resistivity $\rho_{\rm ff}
\sim\rho_n B/H_{c2} < \rho_n$.
\begin{figure}[b]
\includegraphics[width=6cm]{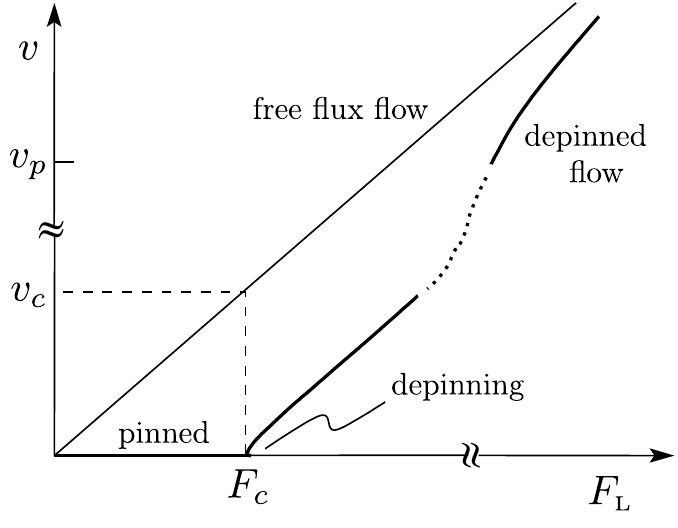}
\caption{Sketch of the force--velocity (equivalent to the current--voltage)
characteristic of a generic type II superconductor in the absence of thermal
fluctuations. For an external force density $F_{\rm \scriptscriptstyle L}$
smaller than the critical force-density $F_c$, the vortex system remains
pinned and $v=0$. For $F_{\rm \scriptscriptstyle L}>F_c$, vortices are
depinned and move with finite velocity $v>0$; at large drives $F_{\rm
\scriptscriptstyle L}\gg F_c$, the characteristic approaches the free
flux-flow regime.  For strong pinning in the dilute-pin limit, we find an
excess-force characteristic, with a linear flux-flow branch shifted by $F_c$
extending over a large regime of velocities $v$ beyond $v_c = F_c/\eta$.
The pinning-force density $\langle F_p (v)\rangle$ changes on the velocity
scale $v_p \gg v_c$ and the force--velocity characteristic approaches the free
flux-flow regime only for $v \gg v_p$.  \label{fig:ivsketch}}
\end{figure}

Material defects lead to vortex pinning
\cite{labusch_69,campbell_72,larkin_79}; they transform the Abrikosov lattice
into a disordered phase \cite{larkin_70,blatter_94,brandt_nat_95_00} and
reestablish the superconductor's ability to carry current free of dissipation.
The dissipative force balance equation is augmented by the velocity-dependent
mean pinning-force density $\langle F_p(v)\rangle$, $\eta v =
F_{\rm\scriptscriptstyle L} - \langle F_p (v) \rangle$, entailing important
modifications of the vortex dynamics $v(F_{\rm\scriptscriptstyle L})$: below
the critical force $F_c =  \langle F_p (v=0)\rangle$, vortex motion is
inhibited; this defines the critical current density $j_c = c F_c/B$.  Above
depinning at $F_c$ (or for currents $j > j_c$), vortices start moving across
defects with an average bulk velocity determined by the velocity dependent
pinning-force density $\langle F_p (v)\rangle$.  The linear flux-flow behavior
with its reduced resistivity $\rho_{\rm ff}$ is assumed only at high drives or
velocities. The full force--velocity ($F_{\rm \scriptscriptstyle L}$--$v$)
characteristic of the superconductor, see Fig.\ \ref{fig:ivsketch}, then
characterizes the zero temperature vortex dynamics in a complete way. With the
driving force $F_{\rm \scriptscriptstyle L}$ proportional to the applied
current $I$ and the voltage drop $V$ across the sample proportional to the
vortex velocity $v$, the force--velocity curve is equivalent to the measured
current--voltage (or $I$--$V$) characteristic.  In this paper, we determine
the force--velocity (or current--voltage) characteristic (see Fig.\
\ref{fig:ivsketch}) of a strongly pinned vortex solid in a generic isotropic
type II superconductor and in the absence of thermal fluctuations.

Vortex pinning has originally been studied for strong pinning centers by
Labusch \cite{labusch_69} (see also Ref.\ [\onlinecite{campbell_72}]).  Strong
pins induce bistable states in the flux-line lattice. They act individually
\cite{labusch_69} and the direct summation of pinning forces is nonzero, i.e.,
$j_c \propto n_p$ with $n_p$ the density of defects or pins; collective
effects due to other pins result in small corrections.  If individual pins are
weak, pinning is collective and vortices are only pinned by the joint action
of many pinning centers \cite{larkin_79}; the direct summation of the forces
induced by individual pins averages out to zero and $j_c \propto n_p^2$ for
the simplest case of non-dispersive weak bulk pinning.  The crossover between
the regimes of weak collective and strong pinning is given by the Labusch
criterion \cite{labusch_69} which involves the ratio $\kappa$ between the
steepest force gradient $\partial_x f_p(x)$ (the largest negative curvature)
of the pinning potential and an effective elasticity $\bar{C}$ of the lattice.
Pinning is strong if the pinning-force gradient dominates the elasticity with
$\kappa > 1$. On the other hand, in a very stiff lattice with large elastic
constants, we have $\kappa < 1$ and pinning is weak and necessarily
collective.

While weak collective pinning has been intensely studied during recent times
\cite{blatter_94,brandt_nat_95_00}, the further development of strong pinning
theory has been less dynamic, although some progress has been made
\cite{campbell_78,matsushita_79,larkin_86,ivlev_91,koshelev_11,willa_15,willa_16}.
Recently, the relation between weak collective versus individual strong
pinning has been analyzed within a pinning diagram \cite{blatter_04}
delineating the origin of static critical forces $F_c$ as a function of defect
density $n_p$ and strength $f_p$.  In the present paper, we focus on the
dynamic aspects of strong pinning.

The force--velocity characteristic derives from the dynamical equation for
vortex motion
\begin{equation}\label{eq:eom}
  \eta v = F_{\rm\scriptscriptstyle L}(j)-\langle F_p (v)\rangle.
\end{equation}
The main difficulty with Eq.\ \eqref{eq:eom} is in the determination of the
velocity-dependent average pinning-force density $\langle F_p(v) \rangle$ (we
choose $\langle F_p(v) \rangle$ to be positive).  Within the framework of weak
collective pinning theory, dimensional \cite{larkin_79,blatter_94} or
perturbative \cite{schmid_73,larkin_74} estimates have been made and provide
results on a qualitative level with a focus on either the perturbative regime at
high velocities \cite{schmid_73,larkin_74} or on the universal regime near
depinning \cite{nar-chauve_92_00}. In concentrating on the strong pinning
situation, we study the limit of dilute pins, i.e., a small pin-density $n_p$,
and consider defects which pin at most one vortex line---we call this the
single-pin--single-vortex strong pinning regime. 

The task of finding the force--velocity characteristic involves three steps:
first, we have to slove the dynamical equation of motion for a vortex line
moving along $x$ and crossing the center of a pinning defect. The solution of
this `microscopic' problem provides us with the time-dependent displacement
field $u(t)$ of the moving vortex and the `elementary' pinning force
$f_p[u(t)]$ acting on the vortex line. Second, a proper average over the
instantaneous force $f_p[u(t)]$ provides the average pinning force $-\langle
f_p (v) \rangle$ per pin acting on the vortex system (with the sign
guaranteeing a positive average pinning-force density $\langle F_p(v)
\rangle$).  This force changes on the `microscopic' velocity scale $v_p$ which
depends on the size and strength of the individual pins and on the elastic and
dynamical properties of the vortex system but not on the pin density $n_p$.
The force $\langle f_p (v) \rangle$ involves an average along the drive
direction $x$; a second average over the transverse dimension $y$ is required
in order to find the average pinning-force density $\langle F_p (v) \rangle$.
This average can be cast into the form of a transverse pinning or trapping
length $t_\perp(v)$ within which vortices passing the defect are pinned.
Since the pins act individually in the small pin-density limit, we obtain the
average pinning-force density in the form $\langle F_p (v) \rangle = n_p
(2t_\perp/a_0) \langle f_p (v) \rangle$.  At $v=0$, the value of $\langle F_p
(v) \rangle$ defines the critical force-density $F_c$.  Third, given the
driving force-density (or current density) $F_{\rm\scriptscriptstyle L}$, we
have to solve the dynamical equation (\ref{eq:eom}) for the velocity $v$. This
`macroscopic' problem defines a second velocity $v_c = F_c/\eta \propto n_p$,
the flow velocity of vortices at $F_c$ in the absence of pinning, and hence
the seeked force--velocity characteristic involves both a microscopic ($v_p$)
and a macroscopic ($v_c$) velocity scale.

Since the above scheme essentially describes a one-particle (in fact, one
vortex-line) problem, it can be solved via analytical and numerical methods
and the results obtained are precise, in opposition to the usual estimates
made in weak collective pinning theory.  Furthermore, the result in the dilute
pin limit (i.e., small $n_p$) is simple and generic: Rewriting the dynamical
equation (\ref{eq:eom}) in the form
\begin{equation}\label{eq:fv}
   \frac{F_{\rm\scriptscriptstyle L}}{F_c}= \frac{v}{v_c}
   + \frac{\langle f_p (v/v_p) \rangle}{f_c}
\end{equation}
makes the dependence on the two velocity scales $v_c$ and $v_p$ explicit.
Since $v_c \propto n_p$ involves the pin density $n_p$, we have $v_c \ll v_p$
and the velocity scales separate in the dilute pin limit.  With $\langle f_p
(v/v_p) \rangle \approx f_c$ for velocities $v/v_p \ll 1$, we find a
characteristic that takes the generic form of an $F_c$-shifted linear curve,
\begin{equation}\label{eq:excess_current}
   v \approx \frac{1}{\eta} (F_{\rm \scriptscriptstyle L} - F_c), 
\end{equation}
see Fig.\ \ref{fig:ivsketch}; the free dissipative flow $v = F_{\rm
\scriptscriptstyle L}/\eta$ is approached only at very high velocities $v \gg
v_p \gg v_c$.  Experiments measuring the current--voltage, i.e., $I$--$V$,
characteristic then should observe an excess-current characteristic $V =
R_\mathrm{ff}(I-I_c)$ with $R_\mathrm{ff}$ the flux-flow resistivity; this
type of characteristic has been widely measured in the past
\cite{strnad_kim_64_65,huebener_79,berghuis_93,zeldov_02,pace_04} and its
microscopic derivation is the main purpose and result of this paper.  In doing
so, we prove the analogue of Coulomb's law of dry friction (describing the
motion of a solid body sliding on a dry surface) for the case of strong vortex
pinning in the dilute limit: In Amontons' first and second laws of friction,
the friction force, corresponding to our $F_c$, is given by the product of the
friction coefficient $k$ and the normal force $F_n$, $F_f = k F_n$. Amontons'
third law or Coulomb's law of dry friction tells, that the kinetic friction at
finite velocity is independent of the sliding velocity $v$, $F_f(v) = F_f = k
F_n$. These laws immediately imply a linear excess-force characteristic $v =
(F - F_f)\Theta(F-F_f)/\eta$ for the driven (by the force $F$) body with
viscous ($\eta$) dynamics and subject to a friction force $F_f$.

Besides this simple and generic result for the overall shape of the
force--velocity characteristic, it is interesting to understand the change in
the pinning-force density $\langle F_p (v) \rangle$ with velocity $v$ and the
underlying mechanism for this velocity dependence, i.e., the analogue of the
corrections to Coulomb's law of dry friction. Figure \ref{fig:lorentz} shows
the result for the average force $\langle f_p (v) \rangle$ (carrying the main
velocity-dependence of $\langle F_p (v) \rangle$) generated by a
Lorentzian-shaped pinning potential.  For very strong pinning with $\kappa \gg
1$, we find a smooth decrease of $\langle f_p (v) \rangle$ with increasing
velocity $v$ with three characteristic velocity regions: Starting with large
velocities $v > \kappa v_p$ and using perturbation theory around the flux-flow
solution, one finds that $\langle f_p(v)\rangle \propto 1/\sqrt{v}$.  An
extended intermediate-velocity regime $v_p < v < \kappa v_p$ appears for large
$\kappa$ values where $\langle f_p(v)\rangle \propto 1/v$. This rapid decay is
due to a collapse of the longitudinal pinning or trapping length
$t_\parallel(v)$ from $\sigma \kappa$ to the geometrical pin size $\sigma$
with increasing velocity $v$. Finally, developing a perturbative theory around
the static (pinned) solution, we find a decreasing pinning force $\langle
f_p(v)\rangle -f_c \propto -\sqrt{v}$ at small velocities $v <  v_p$.  This
square-root {\it decrease} in the pinning force at small velocities $v$
entails an interesting feature in the force--velocity characteristic; the
latter exhibits hysteretic behavior with separated jumps \cite{larkin_86} upon
increasing/decreasing the drive $F_{\rm\scriptscriptstyle L}$ across $F_c$.

The pinning force $\langle f_p (v) \rangle$ looks different when pinning is
weak. For $\kappa < 1$, the critical force $f_c = \langle f_p(v=0)\rangle$
vanishes and the dynamical force $\langle f_p (v)\rangle \propto \sqrt{v}$
increases with velocity. This behavior remains valid for moderately strong
pinning with $\kappa\gtrsim 1$, where the critical current $f_c \propto
(\kappa-1)^2$ assumes a finite value \cite{blatter_04} and the small $v$
correction $\langle f_p (v)\rangle -f_c \propto \sqrt{v}$ still increases with
velocity $v$, see Fig.\ \ref{fig:lorentz}. This square-root {\it increase} in
the pinning-force density then leads to a smooth and reversible quadratic
onset of the velocity, $v \propto (F_{\rm\scriptscriptstyle L}-F_c)^2$ in a
narrow region above $F_c$.

The results at small $\kappa \ll 1$ can all be obtained with the help of
perturbation theory which directly addresses the pinning-force density
$\langle F_p (v)\rangle$.  Thereby it turns out that the expression for the
lowest-order correction $\langle F_p^{\scriptscriptstyle (1)} (v)\rangle$ has
a form which is identical to that of weak collective pinning theory, after
proper identification of the pinning-energy correlators. This also implies,
that we can use the single-pin analysis to rederive the weak collective
pinning results for the critical current density $j_c$, a quite remarkable
finding.

In the following, we first (Section\ \ref{sec:formalism}) derive the
expression for the pinning-force density $\langle F_p (v)\rangle$, simplify
the problem to a manageable version of the single-pin--single-vortex
situation, and derive the Labusch criterion separating weak from strong pins.
In Section \ref{sec:static}, we focus on the static solution and discuss the
universal solution at very strong pinning $\kappa \gg 1$. Section
\ref{sec:dynamic} is devoted to the dynamic solution at finite velocities.  In
order to gain an overview on the problem, we first provide numerical results
for the average pinning force $\langle f_p (v)\rangle$ generated by a
Lorentzian shaped pinning potential and identify the various strong pinning
regimes at high, intermediate, and low velocities.  We discuss the various
analytical schemes to deal with the problem, perturbative methods at large and
small velocities and a self-consistent universal solution in between.  A
special discussion is devoted to the transverse pinning or trapping length
$t_\perp(v)$ and its velocity dependence, see Sec.\ \ref{sec:tperp}. In
Section \ref{sec:Fv}, we discuss the excess-force characteristic as obtained
in the dilute pin-density limit.  Section \ref{sec:models} is devoted to a
brief discussion of model potentials, in particular, the (exactly solvable)
parabolic potential which is often used in the context of simulations on
vortex dynamics in pinning landscapes\cite{Olson}.  In Section
\ref{sec:conclusion}, we summarize our results and place them into context. A
brief account on parts of this work has been given in Ref.\
[\onlinecite{thomann_12}].

%
\section{Formalism \label{sec:formalism}}

We assume a random homogeneous distribution of identical defects of density
$n_p$ and shape
\begin{eqnarray}
   \varepsilon_\mathrm{p}({\bf R},z) &=&
   e_p({\bf R}) \,\delta(z),
\label{eq:e_p_L}
\end{eqnarray}
with depth $e_p$ and width $\sigma \sim \xi \ll a_0$ ($\xi$ and $a_0 =
(\Phi_0/B)^{1/2}$ denote the coherence length or vortex core size and the
distance between vortices, respectively). The pinning force is given by the
gradient ${\bf f}_p({\bf R}) = -\nabla_{{\bf \scriptscriptstyle R}}e_p({\bf
R})$ and we denote its maximal amplitude by $f_p$.  Defects which strongly
suppress the superconducting order parameter within a volume $\sim \xi^3$
generate a pinning potential of depth $e_p\sim H_c^2 \xi^3$, see Ref.\
[\onlinecite{willa_16}] for further details; on the other hand, for small
(atomic) defects \cite{thuneberg_84}, the pinning energy is of order $e_p\sim
H_c^2 \xi \sigma_\mathrm{sc}$, with the electronic scattering cross section
$\sigma_\mathrm{sc}$ replacing the larger area $\xi^2$; such defects then are
more likely to be weakly pinning. Below, we will make occasional use of
Lorentzian-shaped pinning potentials \cite{thomann_12} $e_p({\bf R}) = - e_p /
(1+ R^2/2\xi^2)$ as motivated by the (variational) shape of the vortex core
\cite{schmid_66,clem_75} in combination with a point-like defect.

An ensemble of (homogeneously distributed) defects located at positions ${\bf
r}_{i} = ({\bf R}_i,z_i)$ acts on the flux lines at the positions
\cite{indices} $({\bf R}_\mu + {\bf u}_\mu(z,t),z)$ with the pinning-force
density (exploiting self-averaging)
\begin{eqnarray}
   \label{eq:F_pin}
   \langle {\bf F}_p \rangle
   &=&-\frac{1}{N}\sum_\mu^N\int\frac{dz}{L}\,
   \bf{F}_p({\bf r}_\mu,{\bf u}_\mu), \quad \textrm{with} \\
   \nonumber
   {\bf{F}}_p({\bf r}_\mu,{\bf u}_\mu)
   &=& \frac{1}{a_0^2}
   \sum_i {\bf f}_p\bigl[{\bf R}_\mu+{\bf u}_\mu(z,t)-{\bf R}_i\bigr]
   \delta(z-z_i).
\end{eqnarray}
The minus sign in Eq.\ (\ref{eq:F_pin}) derives from our sign convention in
Eq.\ (\ref{eq:eom}) where $\langle F_p\rangle$ acts against the direction of
the drive.  Here, the coordinates ${\bf r}_\mu = ({\bf R}_\mu,z)$ describe an
ideal triangular Abrikosov lattice with density $a_0^{-2} = B/\Phi_0$ that is
fixed in space.  They serve as reference positions for the vortices that move
with velocity ${\bf v}t$. The dynamical displacement field ${\bf u}_\mu(z,t) =
{\bf v} t + {\bf u}_{p,\mu}(z,t)$ then involves two terms, the first
describing their bulk average motion, while the pinning-induced term ${\bf
u}_{p,\mu}(z,t)$ accounts for the vortex deformations away from the ideal
lattice configuration.  This definition of the displacement field differs from
the one used in the static strong-pinning situation in Ref.\
[\onlinecite{blatter_04}], where the displacement field has been measured with
respect to the free asymptotic positions of the vortices.

The dynamical displacement field ${\bf u}_\mu(z,t)$ can be found from the
self-consistent solution of the vortex equation of motion which we write in
integral form,
\begin{eqnarray} \nonumber
   {\bf u}_{\nu}(z,t)
   &=& {\bf v} t + a_0^2\sum_{\mu} \int dz' dt' \,
   \hat{\bf G}({\bf R}_\nu-{\bf R}_\mu, z-z', t-t')\\
   \noalign{\vspace{-5pt}}
   && \qquad\qquad\qquad\qquad\quad
   \times {\bf F}_p[{\bf r}_{\mu}',{\bf u}_\mu(z',t')],
   \label{eq:u_nu}
\end{eqnarray}
with ${\bf r}_\mu' = ({\bf R}_{\mu},z')$.  In the absence of pinning, the
first term accounts for the Lorentz force in Eq.\ (\ref{eq:eom}) generating
the flux-flow velocity ${\bf v} = {\bf F}_{\rm\scriptscriptstyle L}/\eta$; in
the presence of a pinning-force density $\langle {\bf F}_p\rangle$, the
velocity ${\bf v}$ has to be determined self-consistently from the dynamical
equation (\ref{eq:eom}). The dynamical elastic Green's function $\hat{\bf
G}({\bf r},t)$ is given by the Fourier transform of the matrix
\begin{eqnarray} \label{eq:G}
   G_{\alpha\beta}({\bf k},\omega)&=&\frac{K_\alpha K_\beta/K^2}{c_{11}
   K^2+c_{44} k_z^2-i\eta\omega}\\
   && \nonumber \qquad\qquad\qquad
   +\frac{\delta_{\alpha\beta} -K_\alpha
   K_\beta/K^2}{c_{66} K^2+c_{44} k_z^2-i\eta\omega},
\end{eqnarray}
with the dispersive elastic moduli \cite{blatter_94} $c_{11}({\bf k})$
(compression), $c_{44}({\bf k})$ (tilt), and the non-dispersive shear
$c_{66}$, as well as the dissipative dynamical term $-i\eta\omega$.
\begin{figure}[htb]
\includegraphics[width = 7cm]{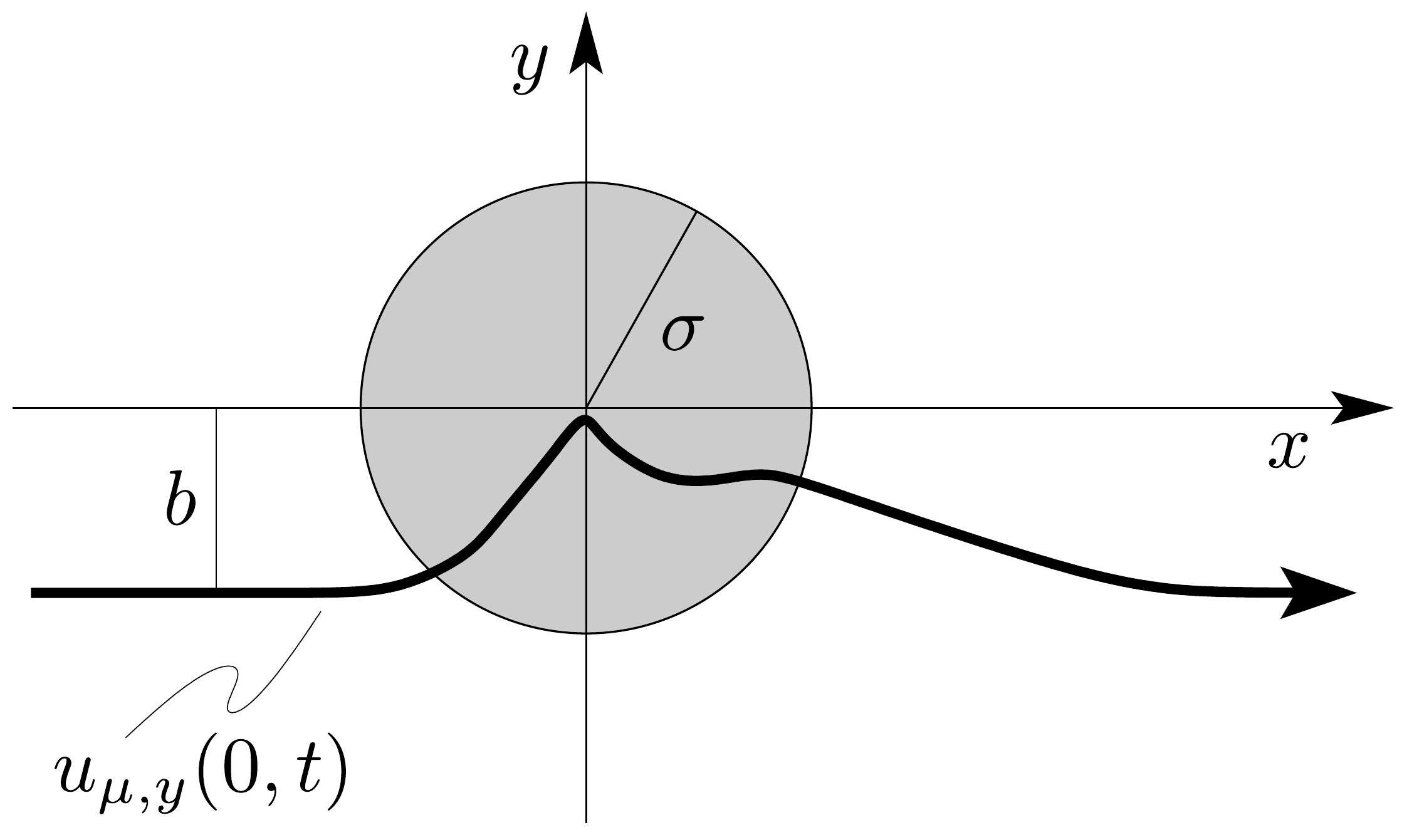}
\caption{\label{fig:impact} Illustration of a vortex trajectory ${\bf u}_\mu
(0,t)$ with a finite impact parameter $u_{\mu,y}(z=0,t=-\infty) = -b$, approaching a
defect with pinning potential of width $\sigma$ located at ${\bf R}_p =(0,0)$.
For a strong pinning center, trapping and depinning are strongly asymmetric, what
gives rise to a finite pinning force $\langle F_p(v)\rangle$.}
\end{figure}

For a dilute density $n_p$ of pinning defects with moderately to strong pinning
forces but trapping no more than one vortex at a time, we can reduce the sum over
$i$ in Eq.\ (\ref{eq:F_pin}) and the sum over $\mu$ in Eq.\ (\ref{eq:u_nu}) to
only one term each; we call this the single-pin--single-vortex limit which
will provide us with results correct to order $n_p$. With the vortex $\mu$
impinging on the defect at $({\bf R_p},z_p)$, we have to find the displacement
field
\begin{eqnarray}
   \nonumber
   u_{\nu,\alpha}(z,t)&=&v_\alpha t
   + \int dt' \, G_{\alpha\beta} ({\bf R}_\nu-{\bf R}_\mu,z-z_p,t-t')\\
   &&\qquad \times
   f_{p,\beta}\bigl[{\bf R}_{\mu}+{\bf u}_{\mu}(z_{p},t')
   -\mbf{R}_p\bigr].
   \label{eq:u_nu_spsv}
\end{eqnarray}
Once we have (self-consistently) solved the dynamical equation for the
displacement field $u_{\mu,\alpha}(z_p,t)$,
\begin{eqnarray} \nonumber
  u_{\mu,\alpha}(z_p,t) &=& v_\alpha t + \int dt'\, G_{\alpha\alpha} (0,t-t') 
  \\ &&\qquad \times 
   f_{p,\alpha}\bigl[{\bf R}_{\mu}+{\bf u}_{\mu}(z_p,t')-{\bf R}_p\bigr],
  \label{eq:u_mu_sc}
\end{eqnarray}
we can find the full displacement field $u_{\nu,\alpha}(z,t)$ by simple
integration of Eq.\ (\ref{eq:u_nu_spsv}).  In Eq.\ (\ref{eq:u_mu_sc}), we have
used that the Green's function $G_{\alpha\beta}(\mbf{r}=0,t)$ is diagonal.

Next, we simplify the expressions for $\langle {\bf F}_p\rangle$ and
$u_{\mu,\alpha}(z_p,t)$, Eqs.\ (\ref{eq:F_pin}) and (\ref{eq:u_mu_sc}), in the
single-pin--single-vortex approximation. We choose a representative vortex at
${\bf R_\mu} = {\bf R}$ and a pin at ${\bf r}_p$ and rewrite the average
pinning-force density Eq.\ (\ref{eq:F_pin}) in the form
\begin{eqnarray}\label{eq:Fpav1}
   \langle {\bf F}_p\rangle &=& -\frac{n_p}{N a_0^2 L} \int \frac{d^2R}{a_0^2}
   \int dz \, \int d^3 r_p \\ \nonumber
   &&\quad \times
   {\bf f}_p\bigl[{\bf R}-{\bf R}_p+{\bf u}({\bf R}-{\bf R}_p,z,t)\bigr] 
   \delta(z-z_p),
\end{eqnarray} 
where we have replaced the sums over $\mu$ and $i$ by the integrations over
$d^2 R/a_0^2$ and $n_p d^3 r_p$. We can choose the pin location $({\bf R}_p,
z_p)$ at the origin and cancel the integral over ${\bf r}_p$ against the
volume $N a_0^2 L$ to arrive at
\begin{eqnarray}\label{eq:Fpav2}
   \langle {\bf F}_p(v)\rangle &=& -n_p \int \frac{d^2R}{a_0^2}\,
   {\bf f}_p\bigl[{\bf R}+{\bf u}({\bf R},0,t)\bigr].
\end{eqnarray}
Furthermore, we note that we can rewrite
the displacement field in Eq.\ (\ref{eq:u_mu_sc}) in the form
${\bf u}({\bf R},0,t) = {\bf v} t + {\bf u}_p({\bf R} +{\bf v}t)$, where
the pinning-induced part ${\bf u}_p$ of the displacement ${\bf u}$
obeys the equation
\begin{eqnarray} \label{eq:u_p_sc}
  u_{p,\alpha}({\bf R}+{\bf v}t) &=& \int dt'\, G_{\alpha\alpha} (0,t-t')
  \\ &&\qquad \times
   f_{p,\alpha}\bigl[{\bf R}+{\bf v}t'+{\bf u}_p({\bf R} +{\bf v}t')\bigr].
  \nonumber 
\end{eqnarray}
The force in Eq.\ (\ref{eq:Fpav2}) can be written as ${\bf f}_p\bigl[{\bf
R}+{\bf v}t +{\bf u}_p({\bf R}+{\bf v}t)\bigr]$ and thus only depends on the
combined argument ${\bf R} + {\bf v}t$, the distance between the vortex and
the pin at time $t$. 

Next, we simplify our task by considering a geometry (see Fig.\
\ref{fig:impact}) with impact parameter $b = 0$, i.e., a vortex hitting the
pin head-on. The average pinning force for this situation can be written as
(see\ Eq.\ \eqref{eq:Fpav2})
\begin{eqnarray} \label{eq:Fpav3}
   \langle f_p(v)\rangle &\approx & -\int \frac{dx}{a_0} \,
   f_p \bigl[x + vt +u_p (x+ vt)\bigr].
\end{eqnarray}
This expression can be further simplified, on the one hand, by selecting
convenient references for the position $x$ and the time $t$, on the other
hand, by choosing between space and time averaging.  Specifically, we change
space to time average, $dx \to v dt$, and then set $x = 0$, what corresponds
to choosing our reference position such that the unperturbed vortex passes the
pin at time $t = 0$. Using $u(vt) = vt + u_p(vt)$, we arrive at the final
result for the average pinning force
\begin{eqnarray}
   \langle f_p(v) \rangle &\approx& - \int \frac{dx}{a_0} f_p \bigl[u(x)\bigr],
\label{eq:f_p}
\end{eqnarray}
where the displacement field $u(x=vt)= u_{\mu,x}(0,t)$ obeys the
self-consistent dynamical equation Eq.\ (\ref{eq:u_mu_sc}) in the simplified
form
\begin{equation}
   u(x) = x + \int_{-\infty}^x \frac{dx'}{v} \, G[0,(x-x')/v] \, f_p[u(x')].
   \label{eq:u}
\end{equation}
In Eq.\ (\ref{eq:u}), we have made use of causality, forcing the Green's
function to vanish at negative times $t-t' < 0$. The coordinate $x=vt$ denotes
the asymptotic position of the vortex at $z=\pm\infty$ (replacing $u$ by $u-x$
produces the displacement field defined in Ref.\ [\onlinecite{blatter_04}])
and we have used the simplified notation $G = G_{xx}$ and $f_p = f_{p,x}$ for
the force along $x$.

In order to find the mean pinning-force density $\langle F_p(v)\rangle$, we
have to perform an additional average over the impact parameter $b = y$ of the
vortex on the pin in Eq.\ (\ref{eq:Fpav2}), see Fig.\ \ref{fig:impact}. This
task is dealt with by equally treating all trajectories within the transverse
trapping range $t_\perp(v)$ of the pin; this can be done exactly in the static
limit \cite{blatter_04}, see below, and is a good approximation at finite
velocities where $t_\perp(v)$ depends on $v$ as discussed in Sec.\
\ref{sec:tperp}.  As a result, the average over $b$ contributes a factor
$2t_\perp(v) /a_0$ and the $y$ component of the force averages to zero.  We
then obtain the final expression for the mean pinning-force density $\langle
F_p(v)\rangle$ along $x$ in the form
\begin{eqnarray}
   \langle F_p(v) \rangle &\approx & n_p \frac{2t_\perp(v)}{a_0} \langle f_p(v)\rangle.
\label{eq:F_p}
\end{eqnarray}
The equations \eqref{eq:f_p}, \eqref{eq:u}, and \eqref{eq:F_p} together with
the dynamical equation (\ref{eq:eom}) define the simplified problem which now
is amenable to a complete (numerical) solution.

The local dynamical Green's function $G(t) \equiv G_{xx}({\bf r} = 0,t)$ is
obtained from the Fourier transform $G_{xx}$ as given by Eq.\ (\ref{eq:G}).
We neglect the compression modes as compared to the softer shear
modes\cite{blatter_94} to find (the average of $K_y^2/K^2$ over the Brillouin
zone leads to the overall factor 1/2 which has been ignored in Ref.\
[\onlinecite{blatter_94}])
\begin{equation}
   G(t)=\Theta(t)\frac{1}{2\eta}\int_\mathrm{BZ} \frac{d^3k}{(2\pi)^3}
   e^{-[c_{66} K^2+c_{44}(\mbf{k})k_z^2]t/\eta}.
   \label{eq:Gt}
\end{equation}
Depending on the time $t$, the integral is either cut by the exponential or by
the Brillouin zone boundary (we use a circularized Brillouin zone with $K_{\rm
\scriptscriptstyle BZ}^2 = 4\pi/a_0^2$); in addition, the dispersive nature of
the tilt modulus has to be accounted for within an intermediate-time regime.

In the static limit, it is the local static elastic Green's function $G({\bf
r} = 0, \omega = 0) = \int_0^\infty dt \, G(0,t)$ which plays an important
role; the latter defines an effective elasticity\cite{willa_16} through
$\bar{C} = 1/G(0,0)$, 
\begin{equation}\label{eq:Cbar}
   \bar{C} \approx  \nu \frac{a_0^2}{\lambda} \sqrt{c_{66}c_{44}}
   \approx 4 \sqrt{\pi}\frac{\varepsilon_0}{a_0},
\end{equation}
where we have used the numerical $\nu = 4$ (but note [\onlinecite{com_nu}]) and
$\varepsilon_0 = (\Phi_0/4\pi\lambda)^2$ denotes the characteristic line
energy of a vortex.

The characteristic time separating different dynamical regimes is given by the
thermal time
\begin{equation}
   t_\mathrm{th}=\frac{\eta}{c_{66} K_{\rm \scriptscriptstyle BZ}^2} \approx
   \frac{4\eta a_0^3}{\sqrt{\pi} \bar{C}},
\label{eq:tth}
\end{equation}
the dissipative relaxation time of short-scale elastic deformations
\cite{blatter_94}.  At long times $t > t_\mathrm{th} (\lambda/a_0)^2$, the
integral in \eqref{eq:Gt} is cut by the exponential at small values of $K$
such that the dispersion in $c_{44}$ can be neglected; with $c_{44}\approx
B^2/4\pi$ and $c_{66}\approx \Phi_0 B/(8\pi \lambda)^2$ one finds the 3D
Green's function \cite{blatter_94}
\begin{eqnarray}
   G^{\rm\scriptscriptstyle 3D}[t_\mathrm{th}(\lambda/a_0)^2 \ll t])
   &\approx&\frac{1}{2\pi}\frac{a_0}{\lambda}
   \frac{1}{\bar{C} t_\mathrm{th}}
   \left(\frac{t_\mathrm{th}}{t}\right)^{3/2}\!\!,
\label{eq:G3d}
\end{eqnarray}
describing a response $G^{\rm \scriptscriptstyle 3D}(t) \propto t^{-d/2}$
involving the entire $d=3$ bulk vortex system.  At intermediate times
$t_\mathrm{th}(\lambda/a_0)^2 > t > t_\mathrm{th}$, the dispersion
$c_{44}\approx B^2/4 \pi \lambda^2 K^2$ in the tilt modulus becomes relevant
and the response is that of a dispersive elastic manifold with an elastic
Green's function behaving as the one of a 4D non-dispersive medium,
\cite{blatter_94}
\begin{eqnarray}
   G^{\rm\scriptscriptstyle 4D}
     [t_\mathrm{th}\ll t \ll t_\mathrm{th}(\lambda/a_0)^2]
   &\approx&\frac{1}{2 \bar C t_\mathrm{th}}
   \left(\frac{t_\mathrm{th}}{t}\right)^{2}\!\!\!.
   \label{eq:G4d}
\end{eqnarray}
For short times $t<t_\mathrm{th}$, the integral is cut by the Brillouin-zone
boundary. This short-time response attains to the dynamics of an individual
vortex line \cite{blatter_94} (in this 1D limit both, longitudinal and
transverse parts of the Green's function Eq.\ \eqref{eq:G}, contribute)
\begin{eqnarray}
   \label{eq:G1d}
   G^{\rm\scriptscriptstyle 1D}
    (t\ll t_\mathrm{th} )
   &\approx&\Theta(t)\frac{2}{\sqrt{\pi}}\frac{1}{\bar C t_\mathrm{th}}
   \left(\frac{t_\mathrm{th}}{t}\right)^{1/2}.
\end{eqnarray}
Note that the time integral in Eq.\ (\ref{eq:u}) is well behaved as the
Green's function is regular (integrable) at long times (since
$G^{\rm\scriptscriptstyle 3D} \propto t^{-3/2}$) as well as at short times
(since $G^{\rm\scriptscriptstyle 1D} \propto t^{-1/2}$), with the main
contribution to the time integral originating from $t_\mathrm{th}$. 

\section{Static solution}\label{sec:static}

The critical pinning-force density $F_c$ is obtained in two steps, where the
first determines the pinning-force average $f_c$ (longitudinal average) and
the second finds the transverse trapping length $t_\perp$ (transverse
average).

\subsection{Longitudinal average $f_c$}\label{sec:long}

\begin{figure}[h]
\includegraphics[width=6.07cm]{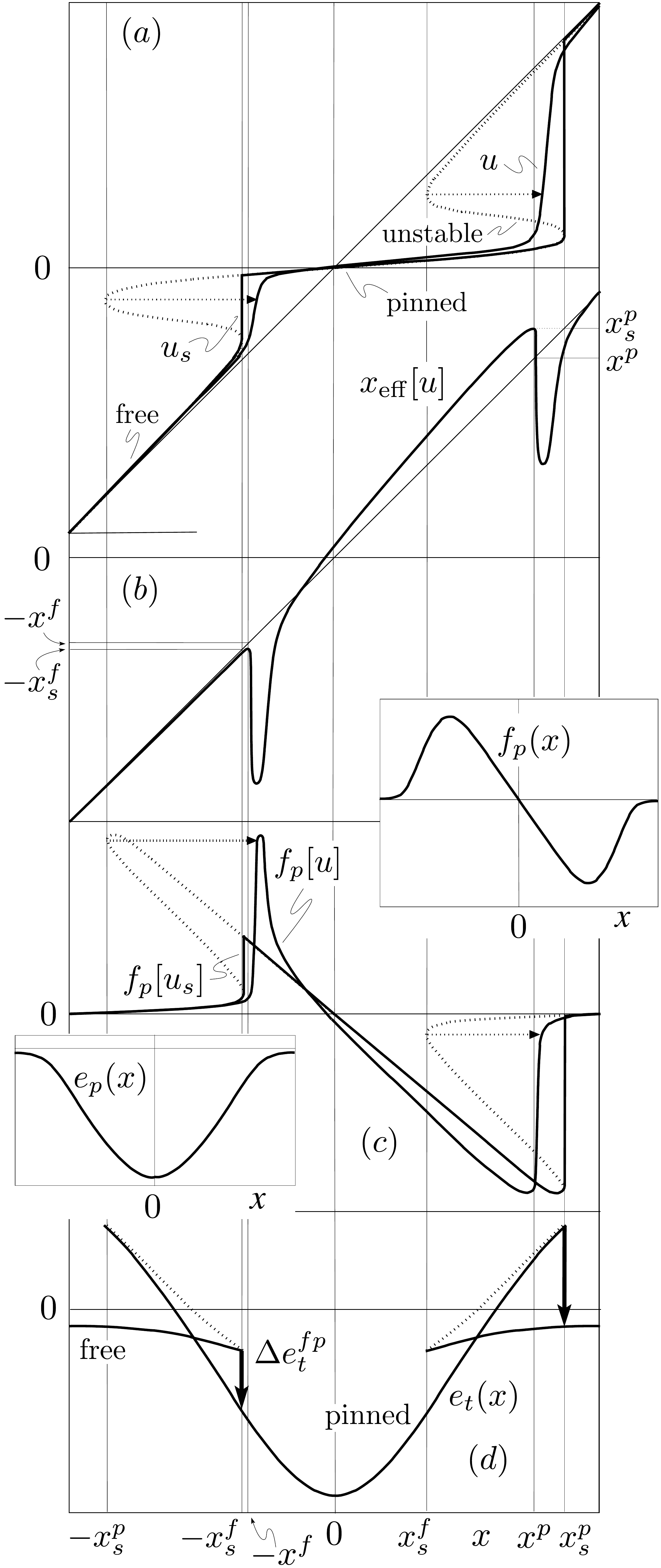}
\caption{Sketch of the static and dynamic displacement fields $u_s(x)$ and
$u(x)$ (a).  The static displacement $u_s(x)$ jumps from the free to the
pinned branch at $-x_s^f$ and back to the free branch at $x_s^p$. In the
smooth dynamic displacement $u(x)$, these jumps are replaced by sharp rises at
the shifted positions $-x^f$ and $x^p$. Below, we will make use of a smooth
multi-valued static solution $\bar{u}_s$ where the jumps are replaced by the
unstable branch $u_s^u$ (dotted).  The dotted arrows refer to the shift
$x_\mathrm{eff}[u](x)$, see (b), that connects the smooth static and dynamic
solutions $\bar{u}_s(x)$ and $u(x)$, respectively.  (b) Effective coordinate
$x_\mathrm{eff}[u](x) = x-\delta x[u](x)$ allowing to express the dynamic
solution through the static one, $u(x) = \bar{u}_s[x_\mathrm{eff}[u](x)]$.
(c) Static and dynamic force profiles $f_p[u_s(x)]$ and $f_p[u(x)]$. (d)
Static total energy profile $e_t(x)$. The insets show the bare force and
energy profiles of the pinning center.} \label{fig:statdyn}
\end{figure}

In the static situation, the self-consistent integral equation (\ref{eq:u})
turns into the simpler algebraic equation
\begin{equation}\label{eq:u_s}
   u_s(x)= x +{f_p[u_s(x)]}/{\bar{C}}.
\end{equation}
Solving Eq.\ \eqref{eq:u_s} self-consistently for $u_s(x)$ and inserting the
result into Eq.\ (\ref{eq:f_p}), we obtain the critical force 
\begin{eqnarray}
   f_c = \langle f_p(v=0)\rangle = - \int \frac{dx}{a_0} \, f_p[u_s(x)].
   \label{eq:Fc}
\end{eqnarray}
The static self-consistency equation \eqref{eq:u_s} can be lifted to a total
energy: we define the total pinning energy $e_t(x)$ as the sum of elastic and
pinning energies, see Fig.\ \ref{fig:statdyn}(d),
\begin{equation}
   e_t(x) \equiv e_t[x;u_s(x)] = \frac{\bar{C}}{2} [u_s(x)-x]^2 + e_p[u_s(x)],
  \label{eq:ex}
\end{equation}
then the total derivative of $e_t(x)$ can be written in the form
\begin{equation} \nonumber
   \frac{d e_t(x)}{dx} = - \bar{C} (u_s-x) + \partial_x u_s[\bar{C} (u_s-x)
    -f_p(u_s)]
\end{equation}
and using Eq.\ \eqref{eq:u_s}, we find that
\begin{eqnarray} 
   f_p[u_s(x)] = - \frac{d e_t(x)}{dx}.
   \label{eq:e'x}
\end{eqnarray}

At weak pinning, the effective static pinning force $f_p[u_s(x)]$ appearing in
Eq.\ (\ref{eq:Fc}) is a single-valued smooth function, resulting in a vanishing
force average 
\begin{eqnarray}
   f_c = \int \frac{dx}{a_0} \frac{d e_t(x)}{dx}
   = \frac{e_t(\infty) - e_t(-\infty)}{a_0} = 0.  \label{eq:Fc0}
\end{eqnarray}
A finite critical force-density $F_c \propto n_p^2$ then is established
through fluctuations in the defect density as described through weak
collective pinning theory.

A strong pin producing a finite average pinning-force density $\propto n_p$ is
characterized by the appearance of bistable solutions (or branches) in the
single-pin problem Eq.\ (\ref{eq:u_s}). The critical force $f_c$ then
depends on the occupation of these branches, with an asymmetric occupation of
the solutions resulting in a finite average force.  Indeed, when typical
values of $f_p/\sigma\bar{C}$ become large, the bare pinning force $f_p(x)$, see
inset of Fig.\ \ref{fig:statdyn}(c), when evaluated at the shifted position
$u_s(x)$, is tilted backward, see Fig.\ \ref{fig:statdyn}(c).  In this
strong-pinning situation the derivative
\begin{equation}
   \partial_x u_s(x)=\frac{1}{1-\partial_u f_p\bigl[u_s(x)\bigr]/\bar{C}}
   \label{eq:u_s'}
\end{equation}
diverges at the positions $\pm x_s^f$ and $\pm x_s^p$ where $\partial_u
f_p[u_s(x)] = \bar{C}$, signalling the appearance of multiple solutions with
pinned ($u_s^p(x)$), unstable ($u_s^u(x)$), and free ($u_s^f(x)$) or unpinned
branches, see Fig.\ \ref{fig:statdyn}. Strong pinning then requires the ratio
\begin{equation}
   \kappa = \max_x \{\partial_u f_p[u_s(x)]\}/\bar{C} = \max_x[f_p'(x)]/\bar{C}
   \label{eq:k}
\end{equation}
to be larger than one, $\kappa >1$; this is the famous Labusch criterion for
strong pinning \cite{labusch_69}.  A vortex incident from the left onto the
defect and moving towards the right will leave the free branch $u_s^f(x)$ at
$-x_s^f$ and jump to the pinned branch $u_s^p(x)$ (see Fig.\
\ref{fig:statdyn}(a), the unstable branch $u_s^u(x)$ and parts of the pinned
branch are jumped over).  After crossing the defect, the vortex will depin
from the pinned branch at $x_s^p$ and jump back to the free branch (the points
$x_s^f$ and $-x_s^p$ are relevant when the vortex moves from right to left).
As a result, the critical pinning force $f_c$ becomes finite and equal to the
sum of energy jumps at $-x_s^f$ and $x_s^p$,
\begin{eqnarray}
   f_c =\! \biggl[ \int_{-\infty}^{-x_s^f}\!\!\!\!\!\!\! + \! \int_{-x_s^f}^{x_s^p}
   \!\!\! +\!\!  \int_{x_s^p}^\infty \biggr] \frac{dx}{a_0}\, \frac{d e_t(x)}{dx} 
   \!=\! \frac{\Delta e_t^{fp}\! + \! \Delta e_t^{pf}}{a_0} 
   \label{eq:fc}
\end{eqnarray}
with the positive jumps $\Delta e_t^{fp}= e_t^f(-x_s^f)-e_t^p(-x_s^f)$ and
$\Delta e_t^{pf} = e_t^p(x_s^p)-e_t^f(x_s^p)$. Hence, it is the asymmetry
between jumping into the pinning well at $-x_s^f$ and out of it at $x_s^p$
which generates the finite (and actually maximal) pinning force
\cite{labusch_69,larkin_79,larkin_86,blatter_04} $f_c$, see Fig.\
\ref{fig:statdyn}(c). Alternatively, Eq.\ (\ref{eq:fc}) may be interpreted in
a (non-equilibrium) statistical sense in terms of an imbalance between the
occupation of the different pinning branches that is produced by the applied
Lorentz force.

\subsection{Trapping lengths $t_\perp$ and $t_\parallel$}\label{sec:trans}

\begin{figure}[ht]
\includegraphics[width = 7cm]{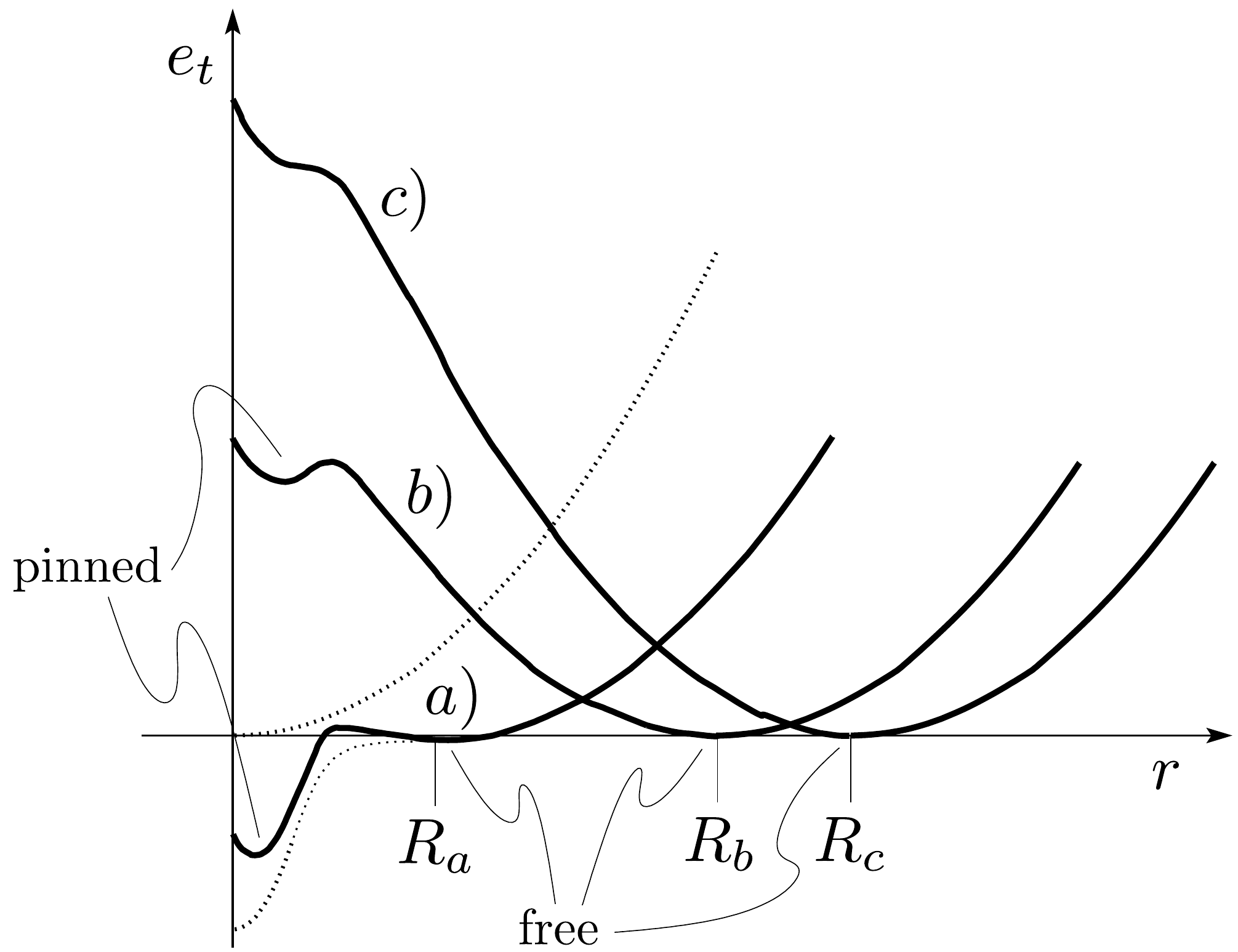}
\caption{\label{fig:eRr} Total energy $e_t(R;r)$ providing the configurational
energy of a vortex at the asymptotic distance $R$ from the pin when its tip
resides at the distance $r$.  Minimizing $e_t(R;r)$ at fixed $R$ with respect
to $r$ defines the static tip position $r_s(R)$. Dotted lines trace the
pinning potential $e_p(r)$ and the elastic energy $\bar{C} r^2/2$.  Solid
lines show the situation for a), $R_s^f < R = R_a < R_s^p$ with two minima,
just before the vortex in the free minimum jumps into the pin at $R_s^f = x_s^f$,
b), $R_s^f < R = R_b < R_s^p$ with two minima just before depinning at $R_s^p =
x_s^p$, and c), $R = R_c > R_s^p$ with one (free) minimum just after depinning
at $R_s^p$.}
\end{figure}

In order to obtain the critical force-density $F_c$, we have to determine the
trapping length $t_\perp$, see Eq.\ \eqref{eq:F_p}. For a radially symmetric
defect potential, this is conveniently done by considering the total energy
$e_t(R;r)$ for a vortex with radial asymptotic and tip positions $R$ and $r$,
see Eq.\ \eqref{eq:ex},
\begin{equation}
   e_t(R;r) = \frac{\bar{C}}{2} (r-R)^2 + e_p(r).
  \label{eq:eRr}
\end{equation}
Plotting this function at fixed $R$ versus $r$, one observes a single (pinned)
minimum in the variable $r$ for $0 < R < R_s^f$, two minima (pinned and free)
when $R_s^f < R < R_s^p$, and again a single (free) minimum for $R > R_s^p$,
see Fig.\ \ref{fig:eRr}; these minima determine the (static) tip position
$r_s(R)$ at given asymptotic position $R$; indeed, the condition $\partial_{r}
e_t(R;r) = 0$ at fixed $R$ reproduces Eq.\ \eqref{eq:u_s} in the form $r = R +
f_p(r)/\bar{C}$ and interrelates asymptotic ($R$) and tip ($r$) positions
of the vortex.  The appearance or disappearance of these minima at $R_s^f~( =
x_s^f)$ and $R_s^p~( = x_s^p)$ signals the beginning or ending of the free and
pinned branches. At these points, the second derivative $\partial^2_{r}
e_t(R;r) = 0$ vanishes as well, i.e., the curvatures in the elastic and
pinning term of Eq.\ \eqref{eq:eRr} compensate and hence
\begin{equation}
   \partial_{r} f_p(r)|_{r(R_s^{f,p})} = \bar{C}.
   \label{eq:R_sfp}
\end{equation}
The latter condition actually determines the critical tip positions $r_s^f =
r(R_s^f)$ and $r_s^p = r(R_s^p)$, while the corresponding asymptotic positions
$R_s^f$ and $R_s^p$ are obtained from solving the force balance equation $r =
R + f_p(r)/\bar{C}$.

In the static situation, a vortex approaching the defect gets trapped as soon
as it enters the circle at $R_s^f$: as the free branch ends at $R_s^f$, the vortex
tip falls into the stable minimum at $r(R_s^f)$ which resides on the pinned
branch, see curve a) in figure \ref{fig:eRr}.  Hence, all vortices impacting
the defect within a distance $R_s^f$ will get trapped and we find that
\begin{equation}
   t_\perp = R_s^f = x_s^f,
  \label{eq:tps}
\end{equation}
see also Ref.\ [\onlinecite{blatter_04}]. Similarly, the vortices remain
trapped by the pin until the asymptotic position $R_s^p$ is reached and we
obtain the longitudinal trapping length 
\begin{equation}
   t_\parallel = R_s^p = x_s^p.
  \label{eq:tls}
\end{equation}
Hence, the mathematical objects $\pm x_s^f$ and $\pm x_s^p$ denoting the
positions where the slope $\partial_x u_s(x)$ of the static displacement
diverges determine the physical lengths $t_\perp$ and $t_\parallel$ where the
vortices get and remain trapped, respectively.

Finally, we combine the results Eq.\ \eqref{eq:fc} for the critical pinning
force and Eq.\ \eqref{eq:tps} for the transverse trapping length into the
expression for the critical-force density $F_c$, see Eq.\ \eqref{eq:F_p},
and obtain 
\begin{equation}
  F_c = \frac{2x_s^f}{a_0^2} n_p [\Delta e_t^{fp} + \Delta e_t^{pf}].
  \label{eq:Fc1}
\end{equation}

\subsection{Universal static solution for very strong pinning $\kappa \gg 1$}
\label{sec:univ_static_sol}

\begin{figure}[ht]
\includegraphics[width = 7cm]{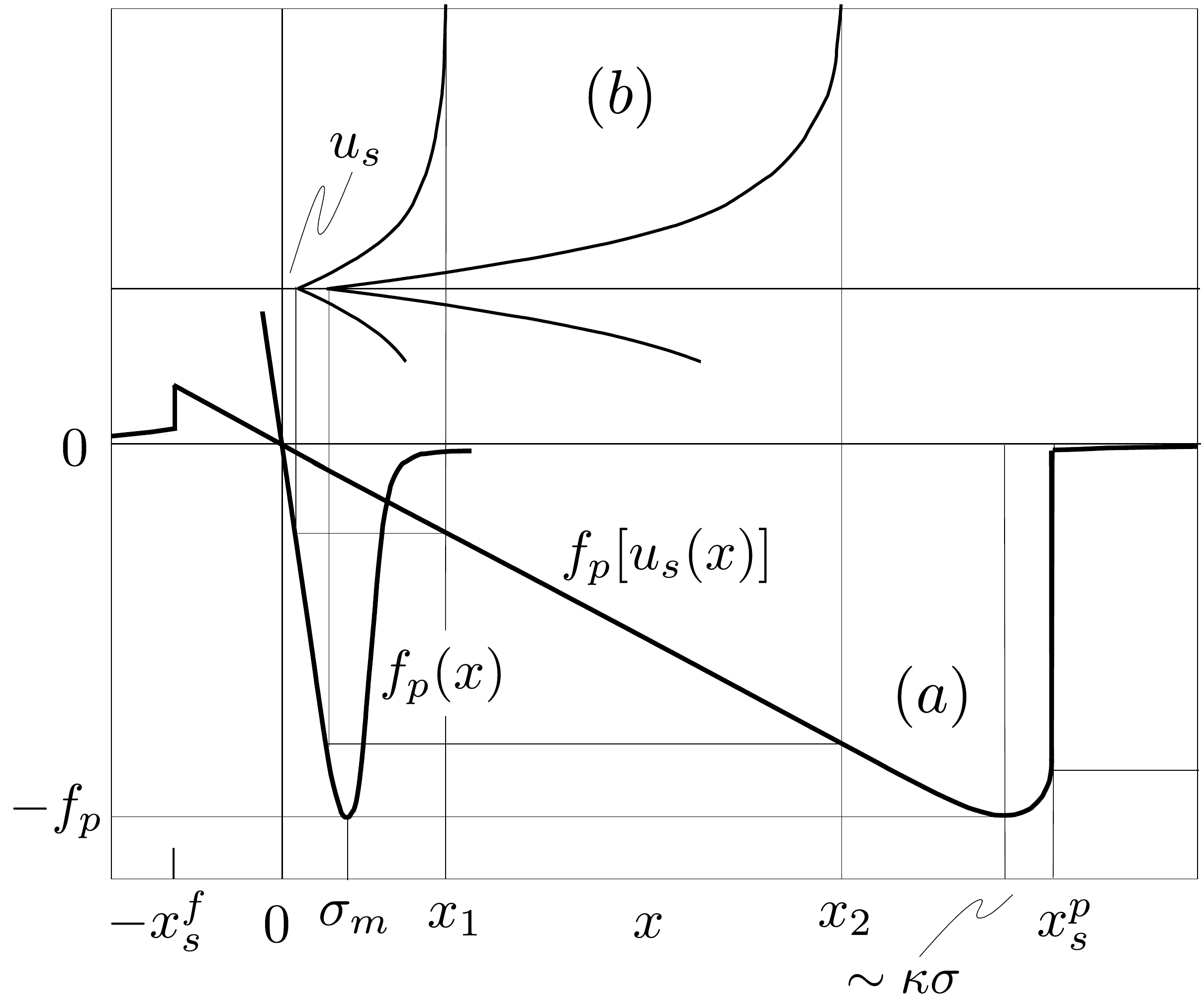}
\caption{\label{fig:sketch_f} (a) Bare and effective static pinning forces
$f_p(x)$ and $f_p[u_s(x)]$ for a strong pinning potential. The steep negative
slope $\partial_x f_p(x)$ on the scale $\sigma$ transforms into the flat
universal slope $-\bar{C} x$ in $f_p[u_s(x)]$ on the larger scale $\kappa
\sigma$.  The rounding before the collapse of $f_p[u_s(x)]$ at $x_s^p$
disappears in the limit $\kappa \to \infty$; the jump $\sim f_p/\kappa$ into
the pin at $-x_s^f$ is small at large $\kappa$.  (b) Vortex deformation for
two asymptotic positions $x_1$ and $x_2$. While $x_2-x_1$ increases on the
scale $\kappa \sigma$ the associated displacements $u_{s1}$ and $u_{s2}$
change on the scale $\sigma$.}
\end{figure}

It turns out, that the above general considerations can be pushed further in
the limit of very strong pinning $\kappa \gg 1$, where a universal solution is
available that is independent of the details of the pinning potential shape.
We start from Eq.\ (\ref{eq:u_s}) by noticing that for the pinned situation,
the last term $f_p[u_s(x)]/\bar{C}$ is large and has to be compensated by the
coordinate $x$, since the tip position $u_s(x)$ on the right has to stay
within the pin and hence is small, $u_s(x) < \sigma$.  As a result, we find
that for very strong pinning, the static force
\begin{equation}
   f_p[u_s(x)] \approx -\bar{C}x
   \label{eq:fu_s_x}
\end{equation}
changes linearly over a wide range until reaching the largest (negative) force
$-\bar{C} x_s^p \approx -f_p$ before depinning, see Fig.\ \ref{fig:sketch_f}.
The latter condition provides an accurate estimate for $x_s^p$ in the very
strong pinning limit,
\begin{equation}
    t_\parallel = x_s^p \approx \frac{f_p}{\bar{C}} \sim \kappa\sigma,
   \label{eq:x_s^p}
\end{equation}
Since $\kappa\sigma$ is large, the residual force after depinning is very
small.  Alternatively, the above result can be found by transforming Eq.\
(\ref{eq:u_s'}) to its force analogue; taking the derivative of Eq.\
(\ref{eq:u_s}) and using Eq.\ (\ref{eq:u_s'}), we find that
\begin{equation}
   \frac{d f_p[u_s(x)]}{dx} = - \frac{\bar{C}}{1-\bar{C}/\partial_u f_p[u_s(x)]}.
   \label{eq:fu_s'}
\end{equation}
Again, for strong pinning, we have $\partial_u f_p[u_s(x)]/\bar{C} \gg 1$ over
a large range $\sim \kappa\sigma$ along the $x$-axis and hence the force
derivative is renormalized to the (constant) effective elasticity $\bar{C}$,
see Fig.\ \ref{fig:sketch_f}.

Next, we discuss the jumps into and out of the pin at $-x_s^f$ and at
$x_s^p$---we will need these results later in the discussion of the small
velocity corrections to $F_c$. We distinguish pins with (long) tails decaying
algebraically with $e_p(r) \propto (\sigma/r)^n$, from compact pins with tails
decaying faster than any power, e.g., $e_p \propto e^{-r/\sigma}$.

The jumps in and out of the pin are determined by the conditions $\partial_u
e_t(x;u) = 0$ and $\partial_u^2 e_p(x;u]) = 0$, see Sec.\ \ref{sec:trans}.
For the jump into the pin, we solve $\partial_u f_p(u) = \bar{C}$, cf.\
Eq.\ \eqref{eq:R_sfp}, and find that $u_s(-x_s^f) \sim - \kappa^{1/(n+2)} \sigma$
for a pin with tails and $u_s(-x_s^f) \sim -\sigma \ln\kappa$ for a compact
pin. The associated asymptotic vortex position $x_s^f$ follows from  Eq.\
\eqref{eq:u_s}; since the jump into the pin takes place at small forces, we
can approximate $-x_s^f \approx u_s(-x_s^f)$ and hence
\begin{equation}
    t_\perp = x_s^f \sim \kappa^{1/(n+2)} \sigma
   \label{eq:x_s^f}
\end{equation}
for a pin with tails. Similarly, for a compact pin, $x_s^f \sim \sigma
\ln\kappa$.  The result for the vortex jumping out of the pin has been found
above, see Eq.\ \eqref{eq:x_s^p}.

The pinning force at $-x_s^f$ can be estimated with the help of Eq.\
\eqref{eq:u_s}: just before the jump, $u_s \approx -x_s^f$ and the force
assumes a small value $f_p(-x_s^f) \approx f_p/ \kappa^{(n+1)/(n+2)}$, while
after the jump, $|u_s| \ll x_s^f$ and hence $f_p(u_s) \approx \bar{C} x_s^f
\sim f_p\kappa^{1/(n+2)-1}$, which is of the same order. For a compact pin,
the force before the jump is $f_p/\kappa$ and assumes a logarithmically larger
value after the jump, $(f_p/\kappa) \ln\kappa$.  
When jumping out of the pin, $u_s(x_s^p)\approx \sigma$ and the pinning force
goes from $-f_p$ before the jump to very small values thereafter, $\sim
-f_p/\kappa^{n+1}$ and $-f_p e^{-\kappa}$ for pins with tails and for compact
pins, respectively. 

The integration over the static force profile $f_p[u_s(x)]$ provides us with a
critical force-density (see Eqs.\ \eqref{eq:Fc} and \eqref{eq:tps}; note
that $f_p[u_s(x)] \approx -\bar{C} x$ on the pinned branch)
\begin{eqnarray}
   \label{eq:F_c_vs}
  F_c &\approx& - n_p \frac{2x_s^f}{a_0^2}
    \int_{-x_s^f}^{x_s^p} dx\, f_p \bigl[u_s(x)\bigr]\\
   &\approx& \Bigl(\frac{x_s^f+ x_s^p}{a_0}\Bigr)^2 n_p x_s^f \bar{C} 
   \approx \frac{\kappa \sigma x_s^f}{a_0^2} n_p f_p,
  \nonumber
\end{eqnarray}
where we have used $x_s^p \sim \kappa \sigma \gg x_s^f$ and $\kappa \sim
f_p/\sigma\bar{C}$ in the last estimate.  The result involves, besides the
maximal force $f_p$ and density $n_p$ of pins, the trapping
area\cite{ivlev_91,blatter_04}
\begin{eqnarray}
   S_\mathrm{trap} = 2 t_\perp t_\parallel \sim 2 x_s^f \, \kappa \sigma.
   \label{eq:S_trap}
\end{eqnarray}
With $\kappa \sim f_p/\sigma \bar{C}$ and $\bar{C}\sim \varepsilon_0/a_0$, see
Eqs.\ \eqref{eq:k} and \eqref{eq:Cbar}, we obtain a field dependence $F_c
\propto B^{(n+1)/2(n+2)}$ for the critical force $F_c$, assuming a pin with
tails.\cite{ivlev_91} For large $n$, this produces the typical strong-pinning
field-dependence $j_c \propto 1/\sqrt{B}$ for the critical current density,
which is cutoff at small fields when bulk 3D strong pinning crosses over to
single-vortex 1D strong pinning at $n_p a_0 \xi^2\kappa \sim
1$.\cite{blatter_04}

\subsubsection{Corrections to the universal static solution}\label{sec:corr}

Let us improve the accuracy of the above analysis and further investigate the
behavior of the vortex near depinning.  In order to find the static force
profile $f_p[u_s(x)]$ near depinning at $x_s^p$, we make use of Eq.\
(\ref{eq:u_s}) and expand the bare potential $f_p(x)$ in the relevant region
around its minimum (near the maximal pinning force pointing along $-x$). We
characterize the minimum in the bare pinning force $f_p(x)$ (i.e., its largest
negative value) through its position $\sigma_m$, the maximum (negative)
pinning force $-f_p$, and the (positive) curvature $f_p'' = f_p/\sigma^2$,
\begin{equation}
   f_p(x) \approx - f_p + \frac{f_p}{2\sigma^2}(x-\sigma_m)^2.
   \label{eq:f_b_x}
\end{equation}
Note that, here, the parameter $\sigma \equiv (f_p''(\sigma_m)/f_p)^{1/2}$ is
a precisely defined model parameter that agrees with the previous (loose)
definition as the pin size for the situation where the defect potential
involves only one length scale.  In order to relate the static displacement
field $u_s$ to the asymptotic coordinate $x$, we combine the above Ansatz for
the bare pinning force $f_p(x)$ with the static self-consistency equation
(\ref{eq:u_s}). Replacing $x \to u_s$ in Eq.\ (\ref{eq:f_b_x}) and using
(\ref{eq:u_s}), $\bar{C} (u_s-x) = f(u_s)$, we find the static displacement
field
\begin{equation}
   u_s(x) \approx \sigma_m + \frac{\sigma}{\bar\kappa}
       -\sqrt{\frac{2\sigma}{\bar\kappa}(x_s^p - x)}
   \label{eq:u_s_x}
\end{equation}
with $\bar\kappa \equiv f_p/\sigma\bar{C}$ and $x^p_s \equiv
\bar\kappa \sigma + \sigma_m +\sigma/2\bar\kappa \approx
\bar\kappa \sigma$ the depinning point, see Fig.\ \ref{fig:sketch_f}.
Making use of Eq.\ (\ref{eq:u_s}) once more, we obtain the static pinning
force
\begin{eqnarray}
   \label{eq:f_u_s_x}
   f_p[u_s(x)] &\approx& -\bar{C} x + f_p \frac{u_s(x)}{\bar\kappa \sigma},
\end{eqnarray}
with the last term a correction of order $\sim f_p /\bar\kappa$.
This force profile decreases linearly with slope
$\bar{C}$ up to $x \approx \bar\kappa \sigma$, reaches its minimum $-f_p$
(the maximum backward force $f_p$) at $x = \bar\kappa \sigma + \sigma_m$ and
increases by $f_p/2{\bar\kappa}^2$ when $x$ increases further by
$\sigma/2\bar\kappa$ in order to diverge upwards at $x =x_s^p$, see
Fig.\ \ref{fig:sketch_f}.

In summary, at very strong pinning (by a defect with tails), we find a
universal static solution where the vortex jumps into the pin at $- x_s^f \sim
-\kappa^{1/(n+2)} \sigma$ and then is deformed linearly in $x$, with the
vortex tip remaining trapped in the pin.  The elastic energy of this
deformation balances the effective pinning force $-\bar{C} x$.  The backward
pointing tip remains fixed onto the defect until reaching the largest force
$-f_p$ with the vortex stretched by $\sim\kappa \sigma$, see Fig.\
\ref{fig:sketch_f}.  When the force decreases again in magnitude, the vortex
remains attached to the pin over the short distance $\sim \sigma/2\kappa$ and
then depins with a sharp forward jump in $u_s(x)$ at $x_s^p$, from $u_s \sim
\sigma$ before to $u_s \sim \kappa\sigma$ after depinning.  With this jump,
the vortex tip depins and ends on the free branch where it experiences a small
residual force $f_p(x_s^p) \sim - f_p/\kappa^{n+1} <0$.  For a vortex with a
finite impact parameter and a radially symmetric defect potential, the above
scenario is still valid, with the vortex jumping into and out of the pin at the
radii $R_s^f = x_s^f$ and $R_s^p = x_s^p$. Note that at very strong pinning,
jumping into and out of the pin are very asymmetric processes, with the vortex
jumping into the pin anywhere along the semicircle with radius $R_s^f$, while
it jumps out of the pin in a narrow, forward directed angle, see Fig.\
\ref{fig:impact}. Hence the depinning process (that determines $\langle
f_p(v)\rangle$) does not depend much on the impact parameter $b$.

\subsection{Static solution for moderately strong pinning $\kappa \gtrsim 1$}
\label{sec:univ_static_solcm}

A similarly accurate analysis can be done at moderate
pinning\cite{blatter_04}. Expanding the pinning potential $e_p(x)$ around the
point $\sigma_{mc}$ of maximal negative curvature, $e_p^{''}(x) = -\bar{C}
\kappa + \alpha(x-\sigma_{mc})^2/2$, the transition to weak pinning can be
described within the Landau formulation of a magnetic phase transition and one
finds the result\cite{blatter_04}
\begin{eqnarray}
   \label{eq:F_c_ms}
   F_c &\approx& 18 \frac{x_s^f}{a_0^2} n_p \frac{\bar{C}^2}{e_p^{''''}}
   (\kappa-1)^2 \sim \frac{\sigma x_s^f}{a_0^2} n_p f_p (\kappa -1)^2, 
\end{eqnarray}
with $e_p^{''''}|_{\sigma_{mc}} = \alpha$. In the last relation, we have used
the estimate $\alpha \sim f_p/\sigma^3$ and $\bar{C}^2/\alpha \sim f_p \sigma$
with $\kappa$ close to unity.

\section{Dynamic solution}\label{sec:dynamic}

Once the Lorentz force density $F_{\rm\scriptscriptstyle L}$ in Eq.\
\eqref{eq:eom} increases beyond the critical force $F_c$, the mean velocity
$v$ becomes finite. According to Eq.\ \eqref{eq:u}, the deformation $u(x)$ of
the vortex is determined by the pinning force $f_p[u(x')]$ averaged over
(past) times $t' = x'/v$ and weighted by the local dynamical Green's function
$G(0,(x-x')/v)$.  The average pinning force $\langle f_p (v) \rangle$, Eq.\
\eqref{eq:f_p}, then depends on the mean velocity $v$, starting at $f_c$ for
$v = 0$ and vanishing at large velocities $v$ as the pins only weakly disturb
the fast flow of vortices. Given the time scale $t_\mathrm{th}$ of $G(0,t)$
and the pinning scale $\kappa\sigma$ over which the force $f_p[u(x')]$ remains
finite, we can estimate the pinning velocity
\begin{equation}
   v_p = \frac{\kappa \sigma}{t_\mathrm{th}} \sim  \frac{f_p}{\eta\, a_0^3}
  \label{eq:vp}
\end{equation}
where dynamical effects start to modify $\langle f_p(v)\rangle$.  The last
expression describes the typical velocity scale of a vortex segment of length
$a_0$ (the vortex tip) moving in the pinning potential $e_p(r)$, $\eta_l a_0
\dot{r} \sim f_p$ and $\eta_l = a_0^2 \eta$ the line viscosity of a vortex.

Given the form of the pinning potential $e_p(r)$, the straightforward
integration of the dynamical equation \eqref{eq:u} gives us access to the
average pinning force $\langle f_p (v) \rangle$ for any velocity $v$ and the
calculation of $t_\perp(v)$ provides us with the pinning-force density
$\langle F_p (v) \rangle$ via Eq.\ \eqref{eq:F_p}. Analytic results, instead,
have to be obtained using different approaches that depend on the velocity
$v$: At high velocities, we use perturbation theory away from flux flow and
directly address the force density $\langle F_p (v) \rangle$, in the
intermediate velocity regime that is present at large values of $\kappa$, we
determine the average force $\langle f_p (v) \rangle$ via construction of a
self-consistent solution for $f_p[u(x)]$, while at low velocities, we find
again the average $\langle f_p (v) \rangle$ using a perturbative approach,
this time away from the static solution. The intermediate and small velocity
results then have to be completed with a calculation of $t_\perp(v)$.

\subsection{Overview on pinning-force averages}\label{sec:regimes}

\begin{figure}[t]
\includegraphics{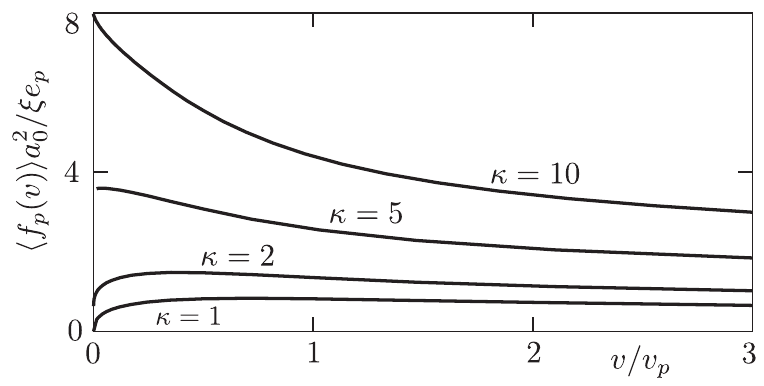}
\caption{\label{fig:lorentz} Velocity-dependence of the pinning-force density
$\langle f_p (v) \rangle$ for Lorentzian pins $e_p(r) = -e_p/(1+r^2/2\xi^2)$
with Labusch parameters ranging from $\kappa=1$ to $\kappa=10$; the typical
depinning velocity scale $v_p$ where the pinning force changes depends on the
Labusch parameter $\kappa$. We haven chosen a field $B = \Phi_0/\lambda^2$,
$a_0 = \lambda$, such that the elastic moduli are non-dispersive. While the
pinning force decreases monotonically for large $\kappa$, it first increases
at small velocities for $\kappa\gtrsim 1$ before eventually decreasing at
large velocities. Original figure published in Ref.\ [\onlinecite{thomann_12}].}
\end{figure}

We first present the results obtained from a numerical forward integration of
Eq.\ (\ref{eq:u}) for a Lorentzian shaped pinning potential (we assume
non-dispersive moduli corresponding to a field $B \sim \Phi_0/\lambda^2$). In
Fig.\ \ref{fig:lorentz}, we show the scaled average pinning force $\langle
f_p(v)\rangle a_0^2/e_p\xi$ versus the scaled velocity $v/v_p$.
Varying the pinning energy $e_p$ at fixed size $\sigma$, we follow the
evolution of the average pinning force $\langle f_p (v) \rangle$ from very
strong pinning $\kappa =10$ to moderately strong pinning at $\kappa \gtrsim
1$. With $e_p \sim H_c^2 \xi^3 \sim \varepsilon_0 \xi$ ($H_c$ the
thermodynamic critical field) the Labusch parameter can naturally access large
numbers $\kappa \sim f_p/\xi\bar{C} \sim (e_p/\xi \varepsilon_0) (a_0/\xi)
\sim a_0/\xi \gg 1$.  For very strong pinning, the critical force $f_c =
\langle f_p (v=0) \rangle$ is large and the pinning force $\langle f_p (v)
\rangle$ decreases when vortices start moving. On approaching the Labusch
point $\kappa = 1$ and for weak pinning ($\kappa < 1$, not shown) the critical
force vanishes and $\langle f_p (v) \rangle$ increases with $v$; this increase
is trivially understood as the pinning force cannot turn negative.  The
vanishing of $f_c$ on approaching $\kappa = 1$ follows a quadratic
behavior\cite{labusch_69,blatter_04}, $f_c \propto (\kappa -1)^2$, see
\eqref{eq:F_c_ms}.

In order to understand the rough functional form of $\langle f_p (v) \rangle$
at small and large velocities $v$, we start from Eq.\ \eqref{eq:f_p} and
expand it about the static ($u_s(x)$) and dynamic ($u(x) \approx x = vt$)
solutions of Eq.\ \eqref{eq:u}, respectively. In the static limit $v=0$,
$f_p[u_s(x')]$ can be taken out of the integral in Eq.\ \eqref{eq:u}; the
remaining integral draws its main contribution from times $t \sim
t_\mathrm{th}$, $\int dt \, G \sim t_\mathrm{th}\, G^{\rm\scriptscriptstyle
1D}(t_\mathrm{th}) \sim 1/\bar{C}$, see Eq.\ \eqref{eq:G1d}. The velocity
correction at small $v$ then derives from cutting this time integral at large
but finite times $t_v \sim \kappa \sigma/v$, as the pinning force $f_p[u(x' =
v t')]$ vanishes when the vortex leaves the pin at $t_v$.  For $t_v >
t_\mathrm{th}$ or $v < v_p$, the integral still picks up its main contribution
near $t_\mathrm{th}$ that produces the static displacement $u_s(x)$; the
correction $\int_{t_v}^\infty d t\, G^{\rm\scriptscriptstyle 3D} (t) \sim
\sqrt{v t_\mathrm{th} / \kappa \sigma}/\bar{C}$ scales with $\sqrt{v/v_p}$ and
hence $\langle f_p (v) \rangle-f_c \propto \sqrt{v/v_p}$ at small velocities.
Note, that a more refined discussion (see Sec.\ \ref{subsec:static_pt} below)
is required in order to explain the sign change in the derivative of $\langle
f_p (v \to 0) \rangle$ with decreasing $\kappa$, see Fig.\ \ref{fig:lorentz}.

In the limit of very high velocity $v$, the displacement $u(x) \approx v t$
and the force \eqref{eq:f_p} vanishes since $\int dx f_p(x) = 0$. Corrections
derive from the second term in Eq.\ (\ref{eq:u}), where the time integral now
is cut on $t_v \sim \sigma/v \ll t_\mathrm{th}$. This implies that the entire
pinning force derives from the integral at short times, $\int_0^{t_v} d t\,
G^{\scriptscriptstyle 1D} (t) \sim \sqrt{\sigma/v t_\mathrm{th}} /\bar{C}$,
and hence the pinning force vanishes as $\langle f_p (v) \rangle \propto
\sqrt{v_p/v}$, resulting in a monotonic decrease of $\langle f_p (v) \rangle$
for large values of $\kappa$.  For small values of $\kappa$, i.e., $\kappa \to
1$, the vanishing of the critical force $f_c = \langle f_p (v=0) \rangle
\propto (\kappa - 1)^2$ implies first an increase of the pinning-force density
$\langle f_p (v) \rangle -f_c \propto \sqrt{v/v_p}$ at small $v$ which is
later followed by the decrease $\propto \sqrt{v_p/v}$ at large $v$, resulting
in the non-monotonic behavior of $\langle f_p (v) \rangle$ shown in Fig.\
\ref{fig:lorentz}.

In the above very high velocity regime, the pinning-induced correction
$u_p \sim f_p \sqrt{\sigma/v t_\mathrm{th}} /\bar{C} \sim \sigma \sqrt{\kappa
v_p/v}$ should be small, $u_p < \sigma$, and thus $v > \kappa v_p$. This leaves a
large intermediate velocity regime $v_p < v < \kappa v_p$ where neither of the
above perturbative approaches can be applied.  Understanding this intermediate
velocity regime requires a more elaborate analysis of Eq.\ \eqref{eq:u}.
Within this region, the two terms on the right are large and nearly compensate
one another to produce a tip position $u(x) \lesssim \sigma$ within the pin.
Hence, Eq.\ \eqref{eq:u} can be written in the form
\begin{equation}
   -vt \approx \int_{-\infty}^t dt' G(0,t-t') f_p[u(t)],
   \label{eq:u_int_reg}
\end{equation}
where we have replaced space by time coordinates. Using the 1D Green's
function \eqref{eq:G1d}, we estimate the right hand side as
$\sqrt{t/t_\mathrm{th}} f_p[u(t)]/\bar{C}$ and therefore $f_p[u(t)] \sim
-v\sqrt{t\, t_\mathrm{th}} \bar{C}$. The linear static force $f_p[u_s(x)]
\approx -\bar{C} x$ of Eq.\ \eqref{eq:fu_s_x} then transforms into a
square-root dynamic force
\begin{equation}
   f_p[u(x)] \sim -\bar{C}\, \sqrt{x\sigma} \sqrt{v/v_\sigma}.
   \label{eq:fu_d_x}
\end{equation}
The maximal pinning force $-f_p$ is reached at $x \sim \kappa \sigma (v_p/v)$,
reduced by a factor $v_p/v$ with respect to the maximal pinning length $x_s^p
\sim \kappa \sigma$ at vanishing velocity. The above results smoothly
interpolate between those found in the small and large velocity regions at $v
< v_p$ and $v > \kappa v_p$, respectively.  Furthermore, one easily convinces
oneself that the pinning time $t_v \sim \kappa\sigma v_p/v^2 < t_\mathrm{th}$
within this velocity region, justifying the use of the 1D Green's function.
Integrating the pinning force \eqref{eq:fu_d_x} over the reduced pinning
length $t_\parallel \kappa\sigma \sqrt{v_p/v}$, we obtain a decaying average
pinning force $\langle f_p(v)\rangle \propto v_p/v$.
\begin{figure}[h]
\includegraphics{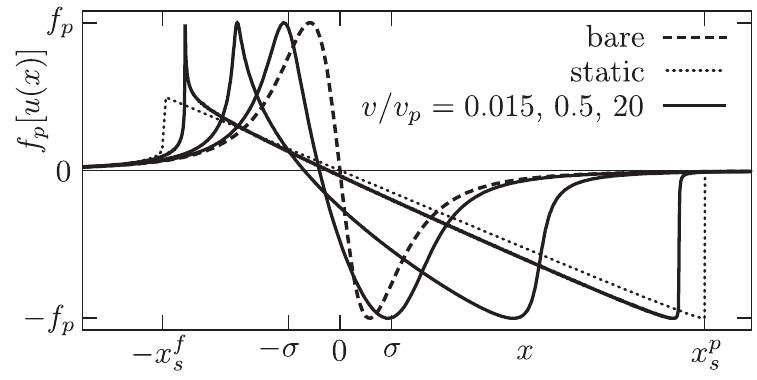}
\caption{\label{fig:lorentz_v} Velocity-dependence of the effective pinning
force $f_p[u(x)]$ for a Lorentzian pin (with maximal force $f_p = 2(3/8)^{3/2}
e_p/\xi$) with Labusch parameter $\kappa=5$ and for velocities $v/v_p=0.015$,
$0.5$, and $20$. The dynamical force profile is compared to the bare force
$f_p(x)$ (dashed) and the static effective force $f_p[u_s(x)]$ (dotted). For
small velocities, the finite-velocity solution follows closely the static
solution $u_s(x)$, motivating an approximative scheme based on the latter as a
starting point.  For large velocities, the force profile approaches that of
the bare force, motivating the use of perturbation theory around free flux
flow.}
\end{figure}

The above results provide a good qualitative understanding of the
velocity-dependence of the effective pinning force $f_p[u(x)]$ shown in Fig.\
\ref{fig:lorentz_v} as calculated for a fixed Labusch parameter $\kappa = 5$
and different velocities $v$.  Indeed, one finds that at low and high
velocities $v$, the effective dynamical force smoothly evolves out of the
static force $f_p[u_s(x)]$ (dotted in Fig.\ \ref{fig:lorentz_v}) and the bare
force $f_p(x)$ (dashed in Fig.\ \ref{fig:lorentz_v}). Furthermore, at
intermediate velocities, the effective pinning force roughly follows a
square-root shape of reduced extent, in agreement with the result
\eqref{eq:fu_d_x}.  The average pinning force $\langle f_p (v) \rangle$ shown
in Fig.\ \ref{fig:lorentz} derives from an average over the local pinning
force $f_p[u(x)]$, see Eq.\ (\ref{eq:F_p}). In the following, we present a
more detailed analysis of the various pinning and velocity regimes.

\subsection{Perturbation theory around flux flow \label{subsec:ffpt}}
When the effect of pinning is small, either at small $\kappa \ll 1$ or at high
velocities $v$, we can use perturbation theory away from flux flow
\cite{schmid_73,larkin_74,blatter_94}. This analysis can be done on the full
two-dimensional problem at $z=0$, using the ansatz ${\bf u}({\bf R},t) = {\bf
v}t +{\bf u}_p({\bf R} + {\bf v}t)$ for the dynamical displacement field with
the pinning contribution ${\bf u}_p$ providing a small correction.  We start
with Eq.\ (\ref{eq:u_p_sc}) and expand the pinning force in ${\bf u}_p$ to
obtain the correction
\begin{equation}
   u_{p,\alpha} ({\bf R}+{\bf v} t) \approx  
   \int_{-\infty}^t dt' G[0,(t-t')] f_{p,\alpha}({\bf R} + {\bf v}t').
   \label{eq:upt}
\end{equation}
Next, we insert this expression back into the formula for the average pinning
force density Eq.\ (\ref{eq:Fpav2}) (we assume a drive along $x$ with ${\bf v} =
(v,0)$ and evaluate the average force at $t=0$) and arrive at
\begin{eqnarray}
   \label{eq:F_pt}
   \langle F_{p}(v) \rangle&\approx &
   -n_p\int \frac{d^2R}{a_0^2}\, \partial_\alpha f_{p,x} ({\bf R})
   \\ \nonumber
   &&\qquad\qquad\quad
   \times \int_0^\infty dt\, G(0,t)\, f_{p,\alpha}({\bf R}-{\bf v}t).
\end{eqnarray}
This result can be brought to the form known from weak collective pinning
theory,
\begin{eqnarray}
   \langle F_{p}(v) \rangle& \approx &
   \int_{0}^{\infty} dt \, G(0,t)\,K^{x\alpha\alpha}({\bf v}t)
   \label{eq:F_pt_K}
\end{eqnarray}
with the pinning energy correlator
\begin{eqnarray}
   K({\bf u})= n_p \int \frac{d^2 R}{a_0^2} 
   \,  e_p({\bf R}) \, e_p({\bf R} - {\bf u})
   \label{eq:Ku}
\end{eqnarray}
and the superscripts in Eq.\ (\ref{eq:F_pt_K}) denoting derivatives with
respect to $u_x$ and $u_\alpha$ (and summation over $\alpha = x,y$).  For weak
pinning, the result Eq.\ (\ref{eq:F_pt_K}) can be used for any velocity $v$.

\subsubsection{Weak pins, small $v$}\label{sec:wpsv}

We first show that the pinning force indeed increases $\propto \sqrt{v}$ at
small velocities. This is easily done in Fourier space, which takes Eq.\
(\ref{eq:F_pt_K}) into the form (we assume a drive along $x$)
\begin{eqnarray}
   \langle F_{p}(v) \rangle&\approx&
    \frac{n_p}{a_0^2}\int \frac{d^2K}{(2\pi)^2}\, 
    K^2 K_x\, |e_p({\bf K})|^2 \\ 
    \nonumber && \qquad  \times \int_0^\infty\!\! dt\, G(0,t) \sin(K_x vt).
   \label{eq:fpt_v_int}
\end{eqnarray}
Using the 3D Green's function Eq.\ (\ref{eq:G3d}) that is relevant at small
velocities $v$, we obtain the result (with the numerical
$\nu^{\scriptscriptstyle <} \approx K(1/2)/(3\sqrt{\pi} \pi^2)$ and $K(m)$ the
complete elliptic integral of the first kind)
\begin{eqnarray}
   \nonumber
   \langle F_{p}(v) \rangle&\approx& \nu^{\scriptscriptstyle <}
    \frac{n_p}{a_0\lambda \bar{C}} 
    \sqrt{\frac{v}{v_\sigma}} 
    \int_0^\infty \!\!\!\! dK\, K^4 \sqrt{K\sigma} |e_p(K)|^2 \\
    \label{eq:fpt_v}
    &\sim& \frac{\sigma^2}{a_0\lambda} n_p \kappa f_p
    \sqrt{\frac{v}{v_\sigma}}, 
\end{eqnarray}
where we provide a simple scaling estimate in the last line.  Here, the
velocity
\begin{eqnarray}
   v_\sigma = \sigma/t_\mathrm{th}
   \label{eq:v_th}
\end{eqnarray}
replaces $v_p$ at weak pinning $\kappa <1$.  As expected, the average pinning
force increases with velocity $v$ from zero with a dependence $\propto
\sqrt{v/v_\sigma}$; this result can be traced back to the time dependence
$\propto t^{-3/2}$ of the 3D Green's function relevant at long times (and
hence small velocities $v$) and a time integral that is cut on $t \sim 1/K_x
v$.

\begin{figure}[h]
\includegraphics{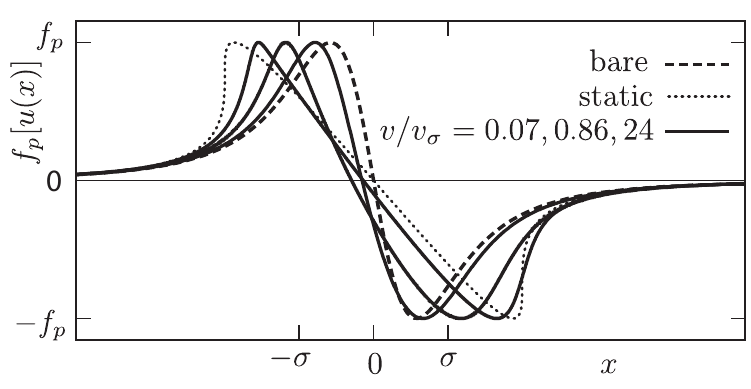}
\caption{\label{fig:wlorentz_v} Velocity-dependence of the effective pinning
force $f_p[u(x)]$ for a Lorentzian pin with Labusch parameter $\kappa= 1.0$
and for velocities $v/v_\sigma=0.07$, $0.86$, and $24$. The dynamical
force profile is compared to the bare force $f_p(x)$ (dashed) and the static
effective force $f_p[u_s(x)]$ (dotted). The asymmetry in the effective pinning
force $f_p[u(x)]$ (narrowing on the left as the vortex is dragged into the
pin, broadening on the right as the vortex is held back in the pin) generates
the finite pinning-force density $\langle F_p (v) \rangle$ at finite velocities,
see Eq.\ (\ref{eq:F_p}).}
\end{figure}

The appearance of a finite pinning-force density for finite velocities $v$ is
illustrated in Fig.\ \ref{fig:wlorentz_v}. In the static situation, the
pinning-force density $\langle F_p(v=0) \rangle=F_c$ vanishes for the single-valued
anti-symmetric solution $u_s(x)=-u_s(-x)$ at weak pinning $\kappa \leq 1$.
For finite velocities, the asymmetric deformation of the effective pinning
force $f_p[u(x)]$ generates a finite positive result $\langle F_p (v) \rangle
> 0$ instead.

The scaling $\propto \sqrt{v/v_\sigma}$ in the pinning-force density persists
even beyond the Labusch point at $\kappa = 1$, the only relevant
change being the appearance of a finite critical force-density
\cite{labusch_69,blatter_04}, $F_c \sim (\sigma x_s^f/a_0^2) n_p f_p
(\kappa-1)^2$, see Eq.\ \eqref{eq:F_c_ms}, while the velocity dependent part
$[\langle F_p (v) \rangle-F_c] \propto \sqrt{v/v_\sigma}$ evolves smoothly
across the Labusch point.

Next, we remain in the weak pinning regime and follow the evolution of the
average pinning-force density $\langle F_p (v) \rangle$ with increasing
velocity. We use the perturbative result for the average pinning-force density
Eq.\ (\ref{eq:F_pt}) in order to find a simple estimate for the evolution of
$\langle F_p (v) \rangle$. The integration over space contributes a factor
$\sigma^2$ due to the pin extension, the force derivative is estimated
as $\partial_x f_p(x) \sim f_p/\sigma$, and the time integral over the Green's
function contributes a factor $tG(0,t)$; the conversion from time to velocity
again involves the pin size $\sigma$, $v \sim \sigma/t$.  Starting from small
velocities $v$, we choose the appropriate Green's function and obtain the
results (with $\kappa < 1$)
\begin{eqnarray}
   \langle F_p (v) \rangle&\sim&
   \frac{\sigma^2}{a_0^2} n_p \kappa f_p 
   \left\{ \begin{array}{ll}
   \displaystyle{ \frac{a_0}{\lambda} \sqrt{\frac{v}{v_\sigma}}, }
      & \displaystyle{\frac{v}{v_\sigma} < \frac{a_0^2}{\lambda^2}, } \\
   \noalign{\vspace{3pt}}
   \displaystyle{ \frac{v}{v_\sigma}, }
      & \displaystyle{\frac{a_0^2}{\lambda^2}<\frac{v}{v_\sigma} < 1,} \\
   \noalign{\vspace{3pt}}
   \displaystyle{ \sqrt{\frac{v_\sigma}{v}}, } 
      & \displaystyle{1 < \frac{v}{v_\sigma}, } \\
           \end{array} \right.
   \label{eq:fpt_all_v}
\end{eqnarray}
with a maximal pinning force $\sim (\sigma/a_0)^2 n_p \kappa f_p$ appearing at
$v_\sigma$.  For weak pinning, all these results remain within the
perturbative regime with $|\delta u| < \sigma$. This is no longer true if we
turn to very strong pinning with $\kappa \gg 1$.

\subsubsection{Very strong pins $\kappa \gg 1$, large $v > \kappa
v_p$}\label{sec:splv}

For very strong pinning $\kappa \gg 1$, we have to make sure that the
displacement $u_p$ remains small, $u_p < \sigma$. A simple estimate
of Eq.\ (\ref{eq:upt}) provides the high-velocity result (we consider short
times and hence use the 1D Green's function)
\begin{equation}
   u_p(x) = \int_0^\infty dt G(0,t) f_p(x-vt) \sim 
                 \sigma \sqrt{\frac{\kappa v_p}{v}}.
   \label{eq:duhv}
\end{equation}
Hence, for very strong pinning, perturbation theory becomes applicable only at
very large velocities $v > \kappa v_p$. We can make use of Eq.\
\eqref{eq:fpt_v_int} and the 1D Green's function Eq.\ \eqref{eq:G1d} to find
the high-velocity result (with the numerical $\nu^{\scriptscriptstyle >}
\approx 2[\Gamma(3/4)]^2/(\sqrt{\pi}\pi^2)$)
\begin{eqnarray}
   \nonumber
   \langle F_{p}(v) \rangle&\approx& \nu^{\scriptscriptstyle >}
    \frac{n_p}{a_0^2 \bar{C}}
    \sqrt{\frac{v_\sigma}{v}} 
    \int_0^\infty \!\!\!\! dK\, (K^4/ \sqrt{K\sigma}) |e_p(K)|^2 \\
   \label{eq:fpt_lv}
    &\sim& \frac{\sigma^2}{a_0^2} n_p f_p
    \sqrt{\frac{\kappa v_p}{v}}.
\end{eqnarray}
This result is consistent with a transverse trapping length $t_\perp \sim
\sigma$, which is reduced by a factor $\kappa^{1/(n+2)}$ with respect to the
static result of Eq.\ \eqref{eq:tps}. As the velocity drops below $\kappa
v_p$, new effects show up which require a self-consistent evaluation of the
vortex dynamics---we will discuss this situation in Section \ref{subsec:scds}
below.

\subsubsection{Weak collective- versus single pins}\label{sec:wcpvsp}

Before turning to the self-consistent dynamical solution, we comment on the
relation between weak collective pinning theory and our single-pin
approximation discussed above. The perturbative result Eq.\ (\ref{eq:F_pt_K})
for the average pinning-force density $\langle F_p (v)\rangle$ obtained within
our single-pin (SP) analysis coincides, to lowest order in $\kappa$ and in the
pin density $n_p$, with the result obtained from weak collective pinning
theory \cite{blatter_94} (WCP). On a first glance, this result may appear as a
surprise, however, the correlations manifest in Eq.\ (\ref{eq:F_pt_K}) arise
quite naturally when constructing a pinning landscape with a finite density
$n_p$ of randomly distributed pins of shape $\varepsilon_p({\bf r})$, see
Eq.\ (\ref{eq:e_p_L}).  The pinning energy density $E_p[{\bf r},{\bf u}]$ of
such a disorder landscape can be written as
\begin{eqnarray}
   E_p[{\bf r},{\bf u}] &=& \sum_{\mu,i} e_p({\bf R}_\mu+{\bf u}_\mu(z)-{\bf R})
   \, \delta^3 ({\bf r}-{\bf r}_i) \nonumber \\
   &=&  \sum_{\mu} e_p({\bf R}_\mu+{\bf u}_\mu(z)-{\bf R}) \, \rho_p({\bf r})
   \label{eq:E_p}
\end{eqnarray}
with the pin density
\begin{equation}
   \rho_p({\bf r})= \sum_i \delta^3 ({\bf r}-{\bf r}_i).
   \label{eq:rho_p}
\end{equation}
The corresponding force field ${\bf F}_p[{\bf r},{\bf u}]$ relates to Eq.\
(\ref{eq:F_pin}) via
\begin{equation}
   \int \frac{d^2 R}{a_0^2} {\bf F}_p[{\bf r},{\bf u}]
   = \sum_\mu {\bf F}_p({\bf r}_\mu,{\bf u}_\mu).
   \label{eq:F_rel_F}
\end{equation}
Assuming {\it self-averaging}, the average pinning-force density $\langle {\bf
F_p} \rangle$ in Eq.\ (\ref{eq:F_pin}) then can be written as the volume
average
\begin{equation}
   \langle {\bf F_p} \rangle = -\frac{1}{V} \int d^3 r\, {\bf F}_p[{\bf r},{\bf u}].
   \label{eq:F_sa}
\end{equation}
Rewriting ${\bf u} = {\bf v} t +{\bf u}_p$ and expanding in ${\bf u}_p$
one obtains (for a drive along $x$)
\begin{eqnarray}\nonumber
   \langle F_{p,x} \rangle &=& -\frac{1}{V} \! \int d^3 r \! \int d^3 r' \! 
   \int^t dt'\, \partial_\alpha\partial_x E_{p}({\bf r},{\bf v}t)
   \\ 
   &&\quad \times G_{\alpha\beta} ({\bf r}-{\bf r}', t-t') \,
   \partial_\beta E_{p}({\bf r}',{\bf v}t').
   \label{eq:F_sa1}
\end{eqnarray}
Inserting the form Eq.\ \eqref{eq:E_p} of the pinning energy landscape and
assuming small and weak pinning defects, the sums over pairs of vortices $\nu$
and $\mu$ and pairs of pins $i$ and $j$ reduce to those terms with $\mu = \nu$
and $i = j$; expressing the sums over vortices and pinning centers as
integrals over $d^2 R/a_0^2$ and $n_p d^3 r_p$, we arrive at
\begin{eqnarray}
   \langle F_{p,x} \rangle &=& -n_p \int^t dt' \, G(0,t-t') 
   \\ \nonumber
   &&\qquad \times \int \frac{d^2R}{a_0^2}\,
   \partial_\alpha\partial_x e_p[{\bf R}+{\bf v}t] \,
   \partial_\alpha e_{p}[{\bf R}+{\bf v}t'],
   \label{eq:F_sa2}
\end{eqnarray}
which, choosing the time $t=0$, is easily reduced to the result Eq.\
(\ref{eq:F_pt_K}). Alternatively, one may take the average over disorder
realizations on the right hand side of Eq.\ (\ref{eq:F_sa1}), represent the 
pinning energy density $E_p$ through Eq.\ \eqref{eq:E_p}, and use
the density-density correlator
\begin{equation}
   \langle \rho({\bf r})\rho({\bf r}') \rangle = n_p^2
   + n_p \delta^3({\bf r}- {\bf r}').
   \label{eq:rhorho}
\end{equation}
The pinning energy correlator then is easily reduced to the usual form
(the reducible term $n_p^2$ in Eq.\ (\ref{eq:rhorho}) is irrelevant in
the present discussion)
\begin{equation}
   \langle  E_{p}({\bf r},{\bf u}) E_{p}({\bf r}',{\bf u}') \rangle
   = \delta^3({\bf r}- {\bf r}') K({\bf u}-{\bf u}')
   \label{eq:E_corr}
\end{equation}
with the $K({\bf u})$ given by Eq.\ (\ref{eq:Ku}) (again, we use that
individual pins are small and weak, such that the sum over vortex pairs
reduces to a single sum over $\mu = \nu$). Hence, the single-pin result Eq.\
(\ref{eq:F_pt_K}) contains all the correlations present in the disorder
landscape described by Eq.\ (\ref{eq:E_p}).

The single-pin approach predicts a vanishing critical force $F_c$ for $\kappa
< 1$, while the weak collective pinning scheme provides a finite result.  This
difference arises due to the different handling of the small velocity limit in
the two approaches: In the weak pinning scenario, we stop decreasing the
velocity $v$ when perturbation theory breaks down as the pinning-induced
correction $\delta v$ becomes of order of the velocity $v$ itself; the
criterion $\delta v \sim v \equiv v_c$ defines a finite (critical)
pinning-force density $F_c^{\rm \scriptscriptstyle WCP} \equiv \langle F_p
(v_c)\rangle$. In the strong pinning scenario, instead, we take $v$ all the
way to zero and obtain a vanishing critical force-density $F_c^{\rm
\scriptscriptstyle SP} = 0$. On the other hand, using the SP result Eq.\
(\ref{eq:F_pt}) and adopting the WCP cutoff scheme, we find a finite critical
current density $j_c$ as well: with the estimate $\langle F_p (v) \rangle^{\rm
\scriptscriptstyle SP} \sim n_p (\sigma/\lambda) (f_p^2/\varepsilon_0)
(v/v_\sigma)^{1/2}$ from Eq.\ (\ref{eq:fpt_all_v}), valid at low velocities,
and the conditions $\langle F_p (v_c) \rangle^{\rm \scriptscriptstyle SP} \sim
\eta v_c \sim j_c B/c$, we obtain the critical current density (with $j_0 \sim
c \varepsilon_0/\Phi_0\xi$ denoting the depairing current density and using
$\sigma \sim \xi$)
\begin{equation}
   \label{eq:jc}
   j_c \sim j_0 (\xi^2/\lambda^2)
   \bigl(n_p a_0^3 f_p^2/\varepsilon_0^2\bigr)^{2} \propto n_p^2,
\end{equation}
in agreement with the results obtained from weak collective pinning theory
\cite{blatter_04}. This result is quite remarkable: first, the critical
current (\ref{eq:jc}) is proportional to $n_p^2$, the {\it square} of the pin
density $n_p$, {\it i.e.}, its origin is in the correlations between pins.
Second, the result is still consistent with the standard SP result $\langle
F_p (v=0) \rangle^{\rm \scriptscriptstyle SP} = 0$, as the latter is an order
$n_p$ result and corrections $\propto n_p^2$ are beyond the standard SP
approach. For strong pinning $\kappa >1$, we already obtain a finite critical
force $\langle F_p (v=0)\rangle^{\rm \scriptscriptstyle SP} \propto n_p$, {\it
linear} in pin density. In this situation, pin-pin correlations are expected
to provide corrections $o(n_p)$ which vanish faster than linear,
allowing us to approach the critical force parametrically closer than in the
WCP situation.

\subsection{Universal self-consistent dynamical solution for very strong pins
with $\kappa \gg 1$} \label{subsec:scds}

Upon decreasing the velocity $v$ below $\kappa v_p$ in the very strong pinning
regime, one has to account for the large deformation of the vortex before
depinning. This is illustrated in Fig.\ \ref{fig:lorentz_v}, where the
effective pinning force extends over a region $t_\parallel$ larger than
$\sigma$ and reaches $x_s^p \sim \kappa \sigma$ at sufficiently low
velocities. Here, we attempt to use Eq.\ \eqref{eq:u} to find a
self-consistent solution for the effective pinning force $f_p[u(x)]$ in the
regime where the vortex deformation is still large.

In the very strong pinning regime and for velocities $v > v_p$ the largest
time scale in the depinning process is given by $t \approx \kappa\sigma/v_p =
t_\mathrm{th}$ and hence the relevant Green's function entering Eq.\
\eqref{eq:u} is given by the 1D expression Eq.\ (\ref{eq:G1d}).  The
self-consistency equation \eqref{eq:u} then can be written as
\begin{eqnarray}
   u(x)-x &=& \frac{1}{\bar{C} \sqrt{v t_\mathrm{th}}}
   \int_{-\infty}^{x} dx'\,\frac{f_p[u(x')]}{\sqrt{x-x'}}.
   \label{eq:sus_0}
\end{eqnarray}
This equation can be solved in three regimes: i) For $x < -\sigma$, the vortex
does not yet feel the pin and the equation is trivially solved by $u(x) = x$.
ii) Similarly, for $x > t_\parallel$, the vortex has depinned and
again $u(x) = x$.  iii) In the intermediate region $-\sigma < x <
t_\parallel \leq \kappa\sigma$, we make use of the Ansatz $f_p[u(x')]
= - \alpha f_p \sqrt{x'+\sigma}$ (see also the discussion leading to Eq.\
\eqref{eq:fu_d_x} in Sec.\ \ref{sec:regimes} above) and find that
\begin{eqnarray}
   x -u(x) &=& \frac{\alpha \, f_p}{\bar{C} \sqrt{v t_\mathrm{th}}}
   \int_{-\sigma}^{x} dx'\,\sqrt{(x'+\sigma)/(x-x')} \\ \nonumber
   &=&
   \frac{\alpha \, \pi \, f_p}{2\bar{C} \sqrt{v t_\mathrm{th}}}(x+\sigma).
   \label{eq:sus_1}
\end{eqnarray}
Dropping the small corrections $u(x) \lesssim \sigma \ll x$ on both sides of
the equation, we find that the constant $\alpha = 2\bar{C} \sqrt{v
t_\mathrm{th}}/\pi f_p \approx 2 \sqrt{v t_\mathrm{th}}/\pi \kappa \sigma$
provides a consistent solution, resulting in the dynamical effective pinning
force
\begin{eqnarray}
   f_p[u(x\gg \sigma)] &\approx& - \frac{2}{\pi} \bar{C}\sqrt{vt_\mathrm{th} x}
   \sim - f_p \sqrt{\frac{v}{v_p}\frac{x}{\kappa\sigma}}
   \label{eq:f_p_int}.
\end{eqnarray}
At depinning, the force assumes its maximal value $-f_p$ and we find the
longitudinal trapping length
\begin{equation}
   \label{eq:seff}
   t_\parallel(v > v_p) \approx  \frac{\pi^2}{4} 
   \Bigl(\frac{f_p}{\bar{C}}\Bigr)^2
   \frac{1}{v t_\mathrm{th}} \sim
   \frac{v_p}{v} \kappa \sigma,
\end{equation}
starting from $\sigma$ at high velocities $v \sim \kappa v_p$ and increasing
to the maximal value $t_\parallel(v_p) \sim \kappa\sigma$ as $v$ drops to
$v_p$, see Sec.\ \ref{sec:static}.  One easily checks that we indeed remain in
the 1D elastic regime throughout this range of velocities: for velocities $v$
larger than $v_p$ one finds that the relevant time $t \sim t_\parallel(v)/v
\sim t_\mathrm{th} (v_p/v)^2 < t_\mathrm{th}$.

Combining Eqs.\ \eqref{eq:f_p_int} and \eqref{eq:seff}, we can provide a good
approximation for the behavior of the dynamical effective pinning force at
large values of $x \gg \sigma$,
\begin{eqnarray}
   f_p[u(x\gg \sigma)] &\approx& - f_p
   \sqrt{x/t_\parallel(v)}
   \label{eq:f_p_int_res}.
\end{eqnarray}
When the vortex depins at $t_\parallel(v)$, it attains its original straight
shape back within a thermal time $t_\mathrm{th}$; this follows from the
dissipative equation of motion for a vortex segment $a_0$ that is displaced by
$u$, $\eta_l a_0 u/ t_\mathrm{dp} \sim \varepsilon_0 u/a_0$ (with $\eta_l =
\eta a_0^2$ the line friction), from which we obtain $t_\mathrm{dp} \sim \eta
a_0^4/\varepsilon_0 \sim t_\mathrm{th}$, see Eq.\ \eqref{eq:tth}. During this
depinning process, the vortex moves by a distance $x_\mathrm{dp} \sim v
t_\mathrm{th} = \sigma v/v_\sigma$. As $v$ increases beyond $v_\sigma$, the
depinning process smoothens, the depinning time is determined by the average
motion, $t_\mathrm{dp} \sim \sigma/v$, and the depinning distance
$x_\mathrm{dp}$ saturates at $\sigma$, 
\begin{eqnarray}\label{eq:xdp}
   x_\mathrm{dp}(v) &\sim&
   \left\{ \begin{array}{ll}
   \displaystyle{\sigma v/v_\sigma, }
      & \displaystyle{v < v_\sigma,} \\
   \noalign{\vspace{3pt}}
   \displaystyle{\sigma, }
      & \displaystyle{v_\sigma < v.} \\
   \end{array} \right.
\end{eqnarray}
These results describe qualitatively well the depinning curves in Fig.\
\ref{fig:lorentz_v}.

Making use of the dynamical effective pinning force \eqref{eq:f_p_int} in the
expression for the average pinning force \eqref{eq:f_p} and cutting the
integral on $t_\parallel(v)$, we find that $\langle f_p (v) \rangle$
decays $\propto v_p/v$ over the extended velocity regime $v_p < v < \kappa
v_p$,
\begin{eqnarray}
   \langle f_p (v) \rangle&\approx& 
   \frac{\pi^2}{6} f_p \Bigl(\frac{f_p}{\bar{C} a_0}\Bigr)^2 
   \frac{a_0}{vt_\mathrm{th}}
   \sim \frac{\kappa\sigma}{a_0} \frac{v_p}{v} f_p,
   \label{eq:fsp_v_int}
\end{eqnarray}
where we have replaced $f_p/\bar{C}\sigma \approx \kappa$ in the second
expression. The pinning-force density $\langle F_p(v)\rangle$ requires
knowledge of the transverse pinning or trapping length $t_\perp(v)$ which will
be calculated in Sec.\ \ref{sec:tperp}.

\subsection{Perturbation theory around static solution}
\label{subsec:static_pt}

At low velocities, the dynamical force $f_p[u(x)]$ remains close to the static
one $f_p[u_s(x)]$, as illustrated in the example of a Lorentzian potential shown
in Fig.\ \ref{fig:lorentz_v}. The basic idea then is to construct a
perturbative analysis away from the static solution $u_s(x)$; the latter rests
on a reformulation of the dynamical equation \eqref{eq:u} that makes use of
the integrated Green's function
\begin{eqnarray}
   G^\uparrow (x) = \int_x^\infty \frac{dx'}{v} G(0, x'/v)
   \label{eq:G^up}
\end{eqnarray}
and rewriting the Green's function $G\bigl[(x-x')/v\bigr]$ in Eq.\
(\ref{eq:u}) through the derivative $-v\partial_x G^\uparrow(x-x')$.
Interchanging the derivative and the integral, we can extract a term
$f_p[u(x)]/\bar{C}$ and arrive at a formula reminiscent of the static variant
(\ref{eq:u_s}),
\begin{equation}
   u(x)= [x-\delta x[u](x)]+\frac{f_p[u(x)]}{\bar{C}},
   \label{eq:dsu}
\end{equation}
with the coordinate shift
\begin{equation}
   \delta x[u](x)=\partial_x \int_{-\infty}^x dx' G^\uparrow(x-x') f_p[u(x')].
\label{eq:d_x}
\end{equation}
The dynamic solution $u(x)$ then can be expressed\cite{larkin_86} through the
smooth multi-valued static solution $\bar{u}_s(x)$ via the self-consistent
coordinate shift $\delta x[u](x)$,
\begin{equation}
   u(x)= \bar{u}_s \bigl[x-\delta x[u](x)\bigr]
    \equiv \bar{u}_s\bigl[x_\mathrm{eff}[u](x)\bigr].
   \label{eq:uus}
\end{equation}
Here, the smooth static solution $\bar{u}_s(x)$ follows the free, unstable,
and pinned branches and thus differs from $u_s(x)$ by a substitution of the
jumps by the unstable branches.  One easily checks that evaluating the static
self-consistency equation \eqref{eq:u_s} at $x-\delta x$ and making use of Eq.\
\eqref{eq:uus} reproduces the dynamical equation \eqref{eq:dsu}.

The integrated Green's function (\ref{eq:G^up}) relevant in the coordinate
shift $\delta x [u] (x)$ assumes the form
\begin{eqnarray}
   \label{eq:Gx}
   G^{\uparrow}(x)\approx
   \left\{ \begin{array}{ll}
   \displaystyle{\frac{1}{\bar{C}}
      ,} & \displaystyle{\frac{x}{v t_\mathrm{th}} < 1, } \\
   \noalign{\vspace{3pt}}
   \displaystyle{\frac{1}{2\bar{C}} \frac{vt_\mathrm{th}}{x},}
      & \displaystyle{1 < \frac{x}{vt_\mathrm{th}} 
            < \frac{\lambda^2}{a_0^2},} \\
   \noalign{\vspace{3pt}}
   \displaystyle{\frac{1}{\pi}\frac{a_0}{\lambda} \frac{1}{\bar{C}}
   \sqrt{\frac{vt_\mathrm{th}}{x}},}
      & \displaystyle{\frac{\lambda^2}{a_0^2}
                      < \frac{x}{vt_\mathrm{th}}.}
           \end{array} \right.
\end{eqnarray}
In the static limit $v\to 0$ the coordinate shift $\delta x[\myu](x)=0$
vanishes, except for two finite spikes of vanishing width $\propto \sqrt{v}$
at $-x_s^f$ and $x_s^p$, and the dynamic displacement approaches the static
one, $u(x) \to u_s(x)$.  Substantial changes in $\delta x[\myu](x)$ (and hence
in $u(x)$ and $\langle f_p(v)\rangle$) are to be expected when the scale $v
t_\mathrm{th}$ of $G^\uparrow(x)$ increases beyond the pinning scale
$\kappa\sigma$ of $f_p[u(x)]$, confirming the estimate $v_p \sim \kappa
\sigma/t_\mathrm{th}$ for the pinning velocity scale in Eq.\ \eqref{eq:vp}.

Given the shape of the displacement $\bar{u}_s(x)$ as discussed above and
illustrated in Fig.\ \ref{fig:statdyn}(a), let us first understand how this
static solution generates the dynamic solution $u(x)$ via the coordinate shift
$\delta x[u](x)$, see \eqref{eq:uus}---the precise understanding of this
mapping is an important step in the construction of the perturbative analysis
of the dynamical displacement $u(x)$ at small velocities $v$ discussed below.

The most noticeable feature in the static solution ${u}_s(x)$ are the jumps at
$-x_s^f$ and $x_s^p$ marking the trapping of the vortex tip and its depinning.
These jumps imply that the vortex does not probe all of the pin potential.  In
the dynamic situation, the vortex tip moves with a finite velocity across the
pin and thus has to pass every point in the pinning potential at some time,
hence the dynamical solution $u(x)$ has to be continuous in $x$.  The relation
(\ref{eq:uus}) maps the continuous function $u(x)$ via the shift function
$\delta x[u](x)$ to the static solution $\bar{u}_s(x)$ of Eq.\ \eqref{eq:u_s},
where $u(x)$ probes all three branches of the static solution $\bar{u}_s(x)$,
free, unstable, and pinned ones. In order to do so, the shift $\delta x[u](x)$
or the effective coordinate $x_\mathrm{eff}[u](x) = x-\delta x[u](x)$ has to
develop (sharp) negative spikes close to the points $-x_s^f$ and $x_s^p$, see
Fig.\ \ref{fig:statdyn}(b). These spikes start at the shifted positions $-x^f$
and $x^p$ defined through the conditions $-x^f -\delta x[u](-x^f) =
x_\mathrm{eff}[u](-x^f) = -x_s^f$ and $x^p -\delta x[u](x^p) =
x_\mathrm{eff}[u](x^p) = x_s^p$ (note that $u(-x^f) = u_s(-x_s^f)$ and $u(x^p)
= u_s(x_s^p)$, see Figs.\ \ref{fig:statdyn}(a) and (b)).  The height of these
spikes are such as to shift the unstable branches $u_s^u(x)$ to the dynamical
solution $u(x)$ (dotted arrows in Fig.\ \ref{fig:statdyn}(a)).  It turns out,
that the shift function $\delta x[u](x)$ is the central quantity in the
perturbative calculation of $\langle f_p (v) \rangle$.

\subsubsection{Very strong pinning with $\kappa \gg 1$}

Our goal is to find the correction in the average pinning force 
\begin{eqnarray}\label{eq:Fuus}
   \langle f_p (v) \rangle-f_c &\equiv& \langle \delta f_p(v)\rangle \\ \nonumber
   &=& -\frac{1}{a_0}\int_{-\infty}^\infty \!\!\!\!
     dx \, \bigl(f_p[u(x)]-f_p[u_s(x)]\bigr).
\end{eqnarray}
The difficulties with the evaluation of Eq.\ \eqref{eq:Fuus} are the steep
slopes in $u(x)$ and jumps in $u_s(x)$ when the vortex enters and leaves the
pin, see Fig.\ \ref{fig:statdyn}. While the jumps in $u_s(x)$ appear at the
positions $-x_s^f$ and $x_s^p$, these positions are shifted to $-x^f$ and
$x_p$ for the dynamical solution $u(x)$. The shifted positions are easily
obtained from Eq.\ (\ref{eq:uus}): The dynamical solution $u(x)$ leaves the
free (pinned) branch of $u_s(x)$ when $x-\delta x[u](x)$ coincides with
$-x_s^f$ ($x_s^p$), hence
\begin{eqnarray}\label{eq:xxs_f}
   -x^f-\delta x[u](-x^f) &=& -x_s^f,
   \\ \label{eq:xxs_p}
   x^p-\delta x[u](x^p) &=& x_s^p.
\end{eqnarray}
At the locations $-x_s^f$, $-x^f$, $x^p$, and $x_s^p$ either the static
($f_p[u_s(x)]$) or dynamic ($f_p[u(x)]$) force changes rapidly between free
and pinned branches. We then can split the integral in Eq.\ (\ref{eq:Fuus})
into appropriate intervals with and without jumps,
\begin{eqnarray}
   \label{eq:xxs_f_split}
   &&\langle \delta f_p (v) \rangle 
   = - \frac{1}{a_0}
   \biggl[
   \int_{-\infty}^{-x_s^f} \!\!\!\!\!\!\!\! dx \, \delta f_{ff}(x)
   + \int_{-x_s^f}^{-x^f}  \!\!\!\!\!\!\!\! dx \, \delta f_{fp}(x)
   \\ \nonumber
   &&
   + \int_{-x^f}^{x^p}  \!\!\!\!\! dx \, \delta f_{pp}(x)
   + \int_{x^p}^{x_s^p}  \!\!\!\! dx \, \delta f_{fp}(x)
   + \int_{x_s^p}^{\infty}  \!\!\!\! dx \, \delta f_{ff}(x)
   \biggr],
\end{eqnarray}
where 
\begin{equation}
   \delta f_{ab}(x) \equiv 
    f_p[u_s^a (x-\delta x[u](x))] -f_p[u_s^b(x)]
\end{equation}
and $u_s^a$ with $a \in \{f,p\}$ denote free and pinned branches of the static
solution $u_s(x)$ (we have expressed $u(x)$ through Eq.\ \eqref{eq:uus} but do
not resolve the steep rise in $u(x)$ on the unstable branch).  Out of the five
terms in Eq.\ (\ref{eq:xxs_f_split}), we can drop the first and last
integrals: For a compact pin, the force is small, of order $f_p/\kappa$ just
before $-x_s^f$ and negligible on the free branch at $x_s^p$, see the
discussion in Sec.\ \ref{sec:univ_static_sol} above. The first integral then
is small by the factor $\kappa^{-3/2}$ as compared to the main terms, while
the last integral is exponentially small.  For a Lorentzian pin with large
tails, the force is of order $f_p/\kappa^{3/4}$ at $-x_s^f$; it contributes a
term smaller by $\kappa^{-9/8}$ as compared to the leading terms. In the last
integral, the force starts at a value of order $-f_p/\kappa^3$ and its
contribution is small by the factor $\kappa^{-6}$.

The second term describes the part of the jump into the pin when the dynamic
solution has not yet responded to the presence of the pin, while the static
solution has already jumped. This term scales the same way as the first
integral: for a compact pin of size $\sigma$, the pinning force $f_p/\kappa$
is still small and the term is small by a factor $\kappa^{-3/2}$, while for a
Lorentzian pin its contribution is small by the factor $\kappa^{-9/8}$ and can
be safely ignored as well.

The important terms are the third and fourth ones describing the change in the
pinning force as the vortex moves through the pin and the force difference
arising from the substantial change in the location of depinning,
respectively,
\begin{eqnarray}
  \label{eq:xxs_f_short}
  \langle \delta f_p (v)\rangle \approx -\frac{1}{a_0}
   \biggl[
   \int_{-x^f}^{x^p}  \!\!\!\!\!\!\!\! dx \, \delta f_{pp}(x)
   +\!\! \int_{x^p}^{x_s^p}  \!\!\!\!\!\!\! dx \, \delta f_{fp}(x)
   \biggr].
\end{eqnarray}
In order to analyze this expression further, we have to investigate the
behavior of the coordinate shift $\delta x[u](x)$, see Eq.\ (\ref{eq:d_x}).
This shift is small, $\propto \sqrt{v}$, over most of the pinning interval
$[-x^f,x^p]$, where the (small) velocity parameter arises from the integrated
Green's function $G^\uparrow(x=vt)$, see the third line of Eq.\ (\ref{eq:Gx}).
A notable exception are the two spikes at $-x^f$ and at $x^p$ where the vortex
jumps into and out of the pin. These spikes are large, of size $x_s^p-x_s^f
\sim \sigma\kappa$, but extend only over a small interval $\sim v
t_\mathrm{th}$, see \eqref{eq:xdp}.

Away from these spikes, we can approximate the coordinate shift $\delta
x[u](x)$ by $\delta x[u_s](x)$.  In fact, the difference $\delta
x[u](x)-\delta x[u_s](x)$ involves the force difference $f_p[u(x)] -
f_p[u_s(x)]$ as it appears in the correction $\langle \delta f_p(v)\rangle$
given by Eq.\ (\ref{eq:Fuus}) above, multiplied with the Green's function
$G^\uparrow (x) \propto \sqrt{v/v_\sigma}$, see Eq.\ (\ref{eq:Gx}).  Taking
the coordinate $x$ through the various regimes, one can show that $\delta
x[u](x) \approx \delta x[u_s](x)$ to order $v$ away from the jumps (with one
factor $\sqrt{v}$ originating from the force difference and another from the
Green's function $G^\uparrow$) and to order $\sqrt{v}$ in a region of order
$\sqrt{v}$ near the jumps (we ignore the region of size $v t_\mathrm{th}$ at
the jumps where the difference is of order $\sigma\kappa$). Furthermore, we
can replace (to precision $\sqrt{v}$) the positions $-x^f$ and $x^p$ by their
static counterparts,
\begin{eqnarray}\label{eq:xsfxf}
   x_s^f-x^f &=& \delta x[u](-x^f) 
   \approx \delta x[u^f_s](-x_s^f) \propto \sqrt{v},
\end{eqnarray}
and similarly (note that we have to make sure that we always stay on
well-defined, continuous branches)
\begin{eqnarray}\label{eq:xspxp}
   x_s^p-x^p &=&- \delta x[u](x^p)
   \approx -\delta x[u^p_s](x_s^p) \propto \sqrt{v}.
\end{eqnarray}

In simplifying Eq.\ (\ref{eq:xxs_f_short}), we make use of the smallness of
$\delta x[u](x)$ in the interval $[-x^f,x^p]$ and expand the integrand $\delta
f_{pp}(x) \approx \partial_x f_p[u^p_s(x)] \, \delta x[u] (x)$.  Next, we
replace (to lowest order in $v$) the unknown dynamical quantities $u(x)$,
$x^f$, and $x^p$ by the known static expressions $u_s(x)$, $x_s^f$, and
$x_s^p$. Finally, we replace the second integral by the product of force
difference $\delta f_{fp}(x) \approx f_p[u_s^f(x_s^p)]- f_p[u_s^p (x_s^p)]$
times the width $x_s^p-x^p \approx -\delta x[u_s^p](x_s^p)$ of the integration
region, exploiting the sharp decay of both $f_p[u_s(x)]$ and $f_p[u(x)]$ at
the boundaries. While the former is a property of the static solution, that
latter is guaranteed by the smallness of the depinning length $x_\mathrm{dp}
\propto v$, see Eq.\ \eqref{eq:xdp}.  We then arrive at the closed expression
for the average pinning force that contains only the static solution $u_s(x)$,
\begin{eqnarray}
  \label{eq:xxs_f_sh_ap}
  \langle \delta f_p (v)\rangle &\approx& \frac{1}{a_0}
   \biggl[
   \int_{-x_s^f}^{x_s^p}  \!\!\!\!\!\!\! dx \, \partial_x f_p[u_s^p(x)]\,
   \delta x[u_s^p](x)\\ \nonumber
   &&
   +\,\delta x[u_s^p](x_s^p)\, \bigl\{f_p[u_s^f(x_s^p)]-f_p[u_s^p(x_s^p)]\bigr\}
   \biggr].
\end{eqnarray}
An equivalent formula was obtained by Larkin and Ovchinnikov in Ref.\
[\onlinecite{larkin_86}] for a periodic pinning model describing large
defects; the regularization introduced in their analysis follows from our
derivation.

Making use of the universal solution for the static strong pinning force in
Sec.\ \ref{sec:univ_static_sol}, we can find the sign of the small-velocity
behavior of $\langle \delta f_p (v)\rangle$. We replace the force gradient by
the (negative) effective elasticity $-\bar{C} \approx - f_p/\kappa\sigma$ in
the first term and set the forces to zero and to $-f_p$ on the free and pinned
branches in the second term to arrive at
\begin{eqnarray}
   \nonumber
   \langle \delta f_p (v)\rangle &\approx& - \frac{f_p}{a_0} \biggl[
   \int_{-x_s^f}^{x_s^p}\!\!\!\!\!\!\!\! dx \frac{\delta x[u_s^p](x)}
    {\kappa\sigma}\! - \! \delta x[u_s^p](x_s^p)\biggr]\\
   \label{eq:xxs_f_simple}
   &=& - \frac{f_p}{a_0} \bigl[\overline{\delta x[u_s^p](x)}
     -\delta x [u_s^p](x_s^p)\bigr],
\end{eqnarray}
where $\>\overline{\cdots}\>$ denotes the average over the interval $[-x_s^f,
x_s^p]$. Finally, we use again the linear form $-\bar{C} x$ for the effective
pinning force and the expression \eqref{eq:Gx}, $G^\uparrow \propto
1/\sqrt{x}$ (valid in the limit $v \to 0$) for the integrated Green's
function, in the calculation of the coordinate shift $\delta x[u_s^p](x)$,
\begin{eqnarray}\nonumber
   \delta x [u_s^p] (x) &\approx& -\bar{C} \partial_x \int_0^x dx'
   G^\uparrow(x-x')\, x'\\ 
   \label{eq:d_x_sp}
   &\sim& -\sigma \frac{a_0}{\lambda} \sqrt{\frac{v\,x}{v_\sigma\,\sigma}}.
\end{eqnarray}
Here, we have ignored the part of $\delta x$ describing the jump into the pin
by starting the integration from $x=0$. Making use of Eq.\ \eqref{eq:d_x_sp}
in \eqref{eq:xxs_f_simple}, the dynamical pinning force correction then
assumes a negative value
\begin{eqnarray}
   \label{eq:d_F_p}
   \langle \delta f_p (v) \rangle 
   \sim -\frac{a_0}{\lambda} f_c \sqrt{\frac{v}{v_p}}.
\end{eqnarray}
This negative correction can be easily understood by noting that $-\delta
x[u_s^p](x)$ is a monotonically increasing positive function and hence the
second (boundary) term in \eqref{eq:xxs_f_simple} always dominates over the
average defining the first term.

The above result applies to the very strong pinning situation with $\kappa \gg
1$.  On the other hand, we have seen in Sec.\ \ref{sec:wpsv} that the
pinning-force density is positive, $\langle F_p(v)\rangle \propto
\sqrt{v/v_\sigma}$ when pinning is weak and $F_c = 0$.  The question arises
about the origin of the sign change in $\langle \delta f_p(v)\rangle$ and for
which value of $\kappa$ this sign change occurs.

\subsubsection{Moderately strong pinning with $\kappa \gtrsim 1$}

In order to understand the sign change in $\langle \delta f_p(v)\rangle$, we
have to be more accurate in our description of $f_p[u_s^p(x)]$ near depinning
at $x_s^p$, as it is the last term in Eq.\ \eqref{eq:xxs_f_simple} that is
strongly modified when $\kappa$ decreases, while the first term remains
unchanged.  Indeed, as $\kappa$ decreases, the decrease $f_p/2\kappa^2$ in the
magnitude of $f_p[u_s^p(x)]$ before the jump at $x_s^p$ increases, see Eq.\
\eqref{eq:f_u_s_x}, leaving a smaller jump at $x_s^p$.  Furthermore, the
coordinate shift $\delta x[u_s^p](x)$ starts decreasing before the jump such
that $x_s^p - x^p$ becomes small. These modifications lead to a reduction of
the second term in Eq.\ (\ref{eq:xxs_f_simple}), which is nothing but the
signature of a decreasing critical force average $f_c$.

More precisely, we can use the result Eq.\ (\ref{eq:f_u_s_x}) for the static
pinning force $f_p[u_s(x)] = f_p[u_s^p(x)]$ to obtain a more accurate
expression for the coordinate shift. We make use of the expression
\eqref{eq:Gx} for the integrated Green's function, $G^{\uparrow} (x-x')
\propto \sqrt{v t_\mathrm{th}/(x-x')}$ for $x-x' > v t_\mathrm{th}$, cut the
integral at $x' = x - v t_\mathrm{th}$, and concentrate on the leading terms
which are large close to $x \approx x_s^p$ to obtain 
\begin{eqnarray}\label{eq:dxus}
   &&\delta x[u](x) \approx \delta x[u_s^p](x) \\ \nonumber
   && \qquad\qquad
   \approx \frac{\partial}{\partial x}\int_{-\infty}^x\!\!\!\!\! dx'
   G^{\uparrow}(x-x')\, f_p[u_s^p(x')]\\ \nonumber
   &&\quad \approx
   -\frac{\sigma}{\pi} \frac{a_0}{\lambda} \sqrt{\frac{v}{v_\sigma}}
   \biggl[2\sqrt{\frac{x}{\sigma}}
   -\sqrt{\frac{\sigma \bar{C}}{2f_p}} 
   \ln\frac{\sqrt{x_s^p}+\sqrt{x}}{\sqrt{x_s^p}
           \!-\!\sqrt{x-v t_\mathrm{th}}}\biggr].
\end{eqnarray}
As expected, the coordinate shift is small in $\sqrt{v}$ and increases in
magnitude $\propto \sqrt{x}$, see Eq.\ (\ref{eq:d_x_sp}).  However, due to the
decrease in the magnitude of the pinning force on approaching $x_s^p$, a
logarithmic correction appears, leading to a sharp collapse of $\delta x$ at
$x_s^p$ which is cutoff (due to the transition to the single-vortex response)
by the term $v t_\mathrm{th}$.  Inserting the result \eqref{eq:dxus} into the
expression \eqref{eq:xxs_f_simple}, we find that the velocity correction to
the average pinning force assumes the form
\begin{eqnarray}\nonumber
   \langle \delta f_p (v) \rangle &\approx& 
   \frac{(x_s^p+x_s^f)}{\pi \lambda} f_p 
   \sqrt{\frac{v t_\mathrm{th}}{x_s^p+x_s^f}} \\ \nonumber
   &&\qquad\qquad\qquad \times \Bigl[-\frac{2}{3} 
   + \frac{1}{4\sqrt{2}}\frac{\sigma\bar{C}}{f_p} \ln\frac{4v_p}{v} \Bigr],\\
   &\sim& \frac{\kappa \sigma}{\lambda} f_p 
   \sqrt{\frac{v}{v_p}} \Bigl[-\frac{2}{3} 
   + \frac{1}{4\sqrt{2}}\frac{\sigma\bar{C}}{f_p} \ln\frac{4v_p}{v} \Bigr].
   \label{eq:Fpv_scale}
\end{eqnarray}
This more accurate analysis shows that $\langle\delta f_p(v)\rangle$ always
increases with $v$ at very small velocities $v$, however, the corresponding
velocity range is exponentially small in $\kappa \sim f_p/\sigma \bar{C}$, $v < v_p
\exp(-\nu \kappa)$ with $\nu = (2 \sqrt{2}/3) 4 \approx 4$, and therefore
irrelevant at very strong pinning with $\kappa \gg 1$.  Indeed, the small
upturn predicted by Eq.\ \eqref{eq:Fpv_scale} is not visible in Fig.\
\ref{fig:lorentz}. The result Eq.\ \eqref{eq:Fpv_scale} straightforwardly
provides the pinning-force density $\langle F_p(v)\rangle$ when combined with
the result for the transverse trapping length $t_\perp(v)$ derived in the next
section.

\subsection{Dynamical transverse trapping length $t_\perp$}
\label{sec:tperp}

The transverse trapping length $t_\perp$ has been found for the two limits of
static pinning at vanishing velocity $v = 0$ and in the perturbative
high-velocity regime $v > \kappa v_p$. The static limit has been discussed in
Sec.\ \ref{sec:trans}, providing us with the result $t_\perp = R_s^f$, the
asymptotic position where the free-branch minimum of the total pinning energy
$e_t(R;r)$ in Eq.\ \eqref{eq:eRr} vanishes; for a very strong pin with tails,
$R_s^f \sim \sigma \kappa^{1/(n+2)}$. This result is parametrically larger
than the one we found at very large velocities $v > \kappa v_p$ using
perturbation theory where $t_\perp \sim \sigma$, see Sec.\ \ref{sec:splv}.
The question then poses itself, how the transverse trapping length shrinks
from $t_\perp \sim \sigma \kappa^{1/(n+2)}$ to $t_\perp \sim \sigma$ as the
velocity $v$ increases.

We start from the static situation and consider the spherically symmetric
total energy of Eq.\ \eqref{eq:eRr} at the critical radius $R_s^f$ where the
first and second derivative vanish, $\partial_r e_t(R;r)|_{R_s^f,r_s^f} = 0$
and $\partial_r^2 e_t(R;r)|_{R_s^f,r_s^f} = 0$, and the free branch
disappears. We consider the limit of very strong pinning (otherwise $t_\perp
\sim \sigma$ follows trivially) and assume a pinning potential with long
tails, $e_p(r) \sim -e_p\,(\sigma/r)^n$, $n=2$ for a Lorentzian shape defect
potential. The critical asymptotic ($R_s^f$) and tip ($r_s^f < R_s^f$)
positions where the free branch terminates then can be estimated to be located
at the radii $R_s^f \gtrsim r_s^f \sim \sigma \kappa^{1/(n+2)}$, where we have
dropped numericals and set $\kappa \sim e_p/\sigma^2 \bar{C}$.  The expansion
of the total potential at $R_s^f$ in the vicinity of $r_s^f$ then is given by
the cubic parabola (we drop a constant of order $\bar{C}\, (r_s^f)^2$)
\begin{equation} \label{eq:cupa}
   e_t(R_s^f; r) = \alpha (r-r_s^f)^3,\quad \alpha \sim \frac{\bar{C}}{r_s^ff}.
\end{equation}
The vortex tip (in the form of a segment of length $a_0$) then follows the
equation of motion $\eta_l a_0 \, \dot{r} = - \partial_r e_t = -\alpha
(r-r_s^f)^2$; in the static limit, every vortex approaching the defect with an
impact parameter $b < R_s^f$ will fall to the center, having an infinity of time
available to overcome the flat potential around  $r_s^f$.  In the dynamical
situation, the vortex passes the pin with a finite velocity $v$ and we have to
include an additional force-term $-\bar{C} [R_s^f - R(t)]$ in the equation of
motion; here, ${\bf R}(t) = (-X^f +vt, Y^f)$ denotes the moving asymptotic
position of the vortex entering the critical radius $R_s^f$ around the defect at
$t = 0$ from the left. Assuming an impact parameter $b$ close to $R_s^f$, we can
expand $R(t) \approx R^f (1-X_s^f vt/ (R_s^f)^2)$, where we follow the trajectory
over a time $t \leq X_s^f/v\equiv t_\mathrm{tra}$ to $X=0$, half the time to
traverse the defect potential. The equation of motion 
\begin{equation} \label{eq:eom_tip}
   \eta_l a_0 \, \dot{r} = -\alpha (r-r_s^f)^2 - \bar{C}(X_s^f/R_s^f) vt
\end{equation}
then picks up an additional force-term that is linear in time $t$ and
independent on position $r$.  Therefore, equation \eqref{eq:eom_tip} can be
brought to the form of Riccati's equation\cite{BenderOrszag}, $\dot{\bar{r}} =
{\bar{r}}^2 + \bar{t}$, where we have introduced the dimensionless variables
$\bar{r}$ and $\bar{t}$ via $r = r_s^f(1- \beta^{1/3} \bar{r})$ and $t =
t_\mathrm{th} \beta^{-1/3}\bar{t}$ with $\beta = (X_s^f/R_s^f) \, v
t_\mathrm{th}/r_s^f$.  Starting from $\bar{r}(\bar{t}=0)=0$, the radius
$\bar{r}$ starts out small and we can drop the $\bar{r}^2$ term to find the
solution $\bar{r}(\bar{t}\,) \approx \bar{t}^{\,2}/2$. When $\bar{r}$ is large, we
can drop the $\bar{t}$ term and obtain the solution $\bar{r}(\bar{t}\,) \approx
1/(\bar{t}^*-\bar{t})$ with $\bar{t}^*$ an integration constant of order unity.  In
order to find the precise location of the divergence, one has to perform a
numerical integration that provides the result $\bar{t}^* \approx 1.986$, see
Ref.\ [\onlinecite{BenderOrszag}].  Thus, as a result, we find that the vortex
tip falls to the center within the time window (we use $\bar{t}^* \approx 2$)
\begin{equation} \label{eq:t_fall}
   t_\mathrm{fall} \lesssim 2 t_\mathrm{th} 
   \Bigl(\frac{r_s^f R_s^f}{X_s^f \, v t_\mathrm{th}} \Bigr)^{1/3}.
\end{equation}
The vortex tip has to fall to the center within a time smaller than
$t_\mathrm{tra} \approx X_s^f/v$, from which we find the condition $(X_s^f)^2
\gtrsim R_s^f \, v t_\mathrm{th}$ (we approximate $r_s^f \approx R_s^f$).  We
thus obtain an upper limit on the impact parameter $b = [(R_s^f)^2 -
(X_s^f)^2]^{1/2}$ of vortices that can be trapped, what provides us with a result
for the transverse trapping length $t_\perp(v)$ at small velocities,
\begin{equation} \label{eq:tperp-cupa}
   R_s^f - t_\perp(v) \sim \frac{v t_\mathrm{th}}{R_s^f}.
\end{equation}

The above analysis applies to impact parameters $b$ close to $R_s^f$ (or small
$X_s^f$) where the tip trajectory is dominated by the slow motion near $r_s^f$. At
small impact parameters $b$ of order a fraction of $R_s^f$, the elastic term in
$e_t$ is no longer relevant and we have to consider the motion of the vortex
tip in the radial pinning potential $e_p(r)$. The equation of motion then has
to be replaced by $\eta_l a_0\, \dot{r} \sim - f_p (\sigma/r)^{n+1}$ and its
integration leads to the trajectory
\begin{equation} \label{eq:eom-ep}
  r(t) = r_0 \bigl(1-t/t_\mathrm{fall}\bigr)^{\! 1/(n+2)}\!\!,
  \quad t_\mathrm{fall} \sim t_\mathrm{th} \Bigl(\frac{r_0}{R_s^f}\Bigr)^{\! n+2},
\end{equation}
where $r_0$ denotes the starting radius at $t = 0$. The fastest fall to the
center appears at the closest approach to the defect and hence we choose $r_0
\sim b$.  Again, the time to fall to the center has to be smaller than the
traversing time which we estimate as $t_\mathrm{tra} \sim b/v$ as given by the
geometry of the problem. As a result, we find that trajectories with an impact
parameter
\begin{equation} \label{eq:tperp-ep}
  b < t_\perp(v) \sim R_s^f \Bigl(\frac{R_s^f}{v t_\mathrm{th}} \Bigr)^{1/(n+1)} 
\end{equation}
fall into the pin. The above result applies for velocities such that $t_\perp
\gtrsim \sigma$, i.e., for $v > v_p$.

Combining the results \eqref{eq:tperp-cupa} and \eqref{eq:tperp-ep}, we find
that the trapping length $t_\perp(v)$ decreases with increasing velocity $v$ on
the velocity scale $R_s^f/t_\mathrm{th} \sim v_p/ \kappa^{(n+1)/(n+2)}$, first
with a correction factor $\propto 1-\kappa^{(n+1)/(n+2)}v/v_p$ and
then with a power law $\propto [v_p/(\kappa^{(n+1)/(n+2)}v)]^{1/(n+1)}$.
Hence, the transverse length $t_\perp(v)$ decreases from $R_s^f$ at $v = 0$ to
$\sigma$ at a velocity $v_p$.

The two trapping lengths $t_\perp(v)$ and $t_\parallel(v)$, see Eq.\
\eqref{eq:seff}, define the trapping area $S_\mathrm{trap}(v) = 2 t_\perp(v)
t_\parallel(v)$ which decreases smoothly from $S_\mathrm{trap} \sim
\kappa^{(n+3)/(n+2)} \sigma^2$ at $v = 0$ to $S_\mathrm{trap} \sim \sigma^2$ at
$v \sim \kappa v_p$ and remains constant thereafter.  Within the region $0< v
< v_p$, this decrease is due to the reduction of $t_\perp$ from
$\kappa^{1/(n+2)}\sigma$ to $\sigma$, while for velocities $v_p < v < \kappa
v_p$ it is the longitudinal length $t_\parallel$ that shrinks from
$\kappa\sigma$ to $\sigma$.

We note that the velocity scale $v_\sigma \kappa ^{1/(n+2)}$ can also been
obtained from a perturbative analysis: for a vortex with asymptotic position
$-x < 0$ passing the pin's center at $t=0$, the pinning-induced correction
\eqref{eq:upt} can be estimated as
\begin{eqnarray} \label{eq:up-pt}
   u_p \sim \frac{\kappa \sigma^{n+2}}{t_\mathrm{th}}
   &\biggl[ \displaystyle{\int_0^{t_\mathrm{th}} \!\!\!\! dt
   \Bigl(\frac{t_\mathrm{th}}{t}\Bigr)^{1/2}  \!\!
   + \!\!\int_{t_\mathrm{th}}^\infty\!\!\!\!  dt \Bigl(\frac{t_\mathrm{th}}{t}
   \Bigr)^{3/2}  \biggr]} \\ \nonumber
   & \times \displaystyle{\frac{x+vt}{[\sigma^2+(x+vt)^2]^{(n+2)/2}}}.
\end{eqnarray}
For small velocities $v < x/t_\mathrm{th}$, the integral assumes its main
contribution from $t \sim t_\mathrm{th}$ and we find that $u_p \sim \kappa
\sigma \, (\sigma/x)^{n+1}$, while for larger velocities the integral is cut
at $v/x$ and $u_p \sim \kappa \sigma \, (\sigma/x)^{n+1}(x/v
t_\mathrm{th})^{1/2}$. The perturbation theory breaks down when $u_p \sim x$,
i.e., at the distance $x_s^f$ at small velocities and at $x_s^f \,
(v_\sigma\kappa^{1/(n+2)} /v)^{1/2(n+3/2)}$ at high velocities, implying a
reduction of the large-distance perturbative region on the velocity scale
$v_\sigma\kappa^{1/(n+2)}$.

\subsection{Summary of force densities}\label{sec:sum_dim_est}

We conclude this section with a summary of results for the dynamic pinning
force density $\langle F_p(v)\rangle ( = F_c + \langle \delta F_p (v)
\rangle)$ on the level of dimensional estimates. For small and intermediate
velocities $v < \kappa v_p$, more accurate expressions are obtained by
combining the results for the average pinning force $\langle f_p(v)\rangle$,
Eqs.\ \eqref{eq:fc}, \eqref{eq:fsp_v_int}, and \eqref{eq:Fpv_scale}, with
those for the transverse trapping length $t_\perp(v)$, Eqs.\
\eqref{eq:tperp-cupa} and \eqref{eq:tperp-ep}, and using Eq.\ \eqref{eq:F_p}.
At high velocities $v > \kappa v_p$, the perturbative result Eq.\
\eqref{eq:fpt_lv} for $\langle F_p(v)\rangle$ can be used.

Starting out at $v=0$, we have the critical force-density
\begin{eqnarray}\label{eq:Fc_sum}
   F_c \sim \frac{\kappa \sigma x_s^f}{a_0^2} n_p f_p \frac{(\kappa-1)^2}{\kappa^2}
\end{eqnarray}
in a form that is valid for all values $\kappa \geq 1$. The corrections at
small velocities $v < v_p$ are given by $\langle \delta F_p(v)\rangle = (2
t_\perp(v)/a_0) n_p \langle \delta f_p(v)\rangle$ with
\begin{eqnarray}
   \langle\delta f_p (v)\rangle &\sim& 
   \left\{ \begin{array}{ll}
   \displaystyle{ - \frac{\kappa\sigma}{\lambda} f_p 
   \Bigl(\frac{v}{v_p}\Bigr)^{\!\frac12},}
   &\displaystyle{\frac{v}{v_p}<\frac{a_0^2}{\lambda^2},}\\
   \noalign{\vspace{5pt}}
   \displaystyle{ - \frac{\kappa\sigma}{a_0} f_p \frac{v}{v_p},}
   &\displaystyle{\frac{a_0^2}{\lambda^2} < \frac{v}{v_p} < 1,}
   \end{array} \right.
   \label{eq:f_sv_pta}
\end{eqnarray}
and
\begin{eqnarray}
   t_\perp (v) &\sim&
   \left\{ \begin{array}{ll}
   \displaystyle{ x_s^f \Bigl(1-\frac{v\kappa^{\frac{n+1}{n+2}}}{v_p}\Bigr),}
   &\displaystyle{\frac{v}{v_p} \ll \kappa^{-\frac{n+1}{n+2}},}\\
   \noalign{\vspace{5pt}}
   \displaystyle{ x_s^f \Bigl(\frac{v_p}{v \kappa^{\frac{n+1}{n+2}}}
   \Bigr)^{\frac{1}{n+1}}\!\!,}
   &\displaystyle{ \kappa^{-\frac{n+1}{n+2}} < \frac{v}{v_p} < 1,} \\
   \noalign{\vspace{5pt}}
   \displaystyle{ \sigma,}
   &\displaystyle{ 1 < \frac{v}{v_p}.}
   \end{array} \right.
   \label{eq:t_perp_sum}
\end{eqnarray}
The corresponding results for moderate values of $\kappa$ remain unchanged up
to a sign change in $\langle \delta f_p(v)\rangle$, i.e., $\langle f_p(v)\rangle$
increases with $v$.  Increasing the velocity beyond $v_p$, we have the following
results for the pinning-force density $\langle F_p(v)\rangle$, see Eqs.\
\eqref{eq:fsp_v_int} and \eqref{eq:fpt_lv},
\begin{eqnarray}
   \langle F_p (v)\rangle \sim
   \frac{\kappa \sigma^2}{a_0^2} n_p f_p
   \left\{ \begin{array}{ll}
   \displaystyle{ \frac{v_p}{v},}
      &\quad
   \displaystyle{1 < \frac{v}{v_p} < \kappa,} \\
   \noalign{\vspace{3pt}}
   \displaystyle{ \sqrt{\frac{v_p}{\kappa v}},}
      &\quad \displaystyle{\kappa < \frac{v}{v_p}. } \\
           \end{array} \right.
   \label{eq:f_sv_ptb}
\end{eqnarray}
Again, these results remain valid as $\kappa$ is reduced towards the Labusch point.
Note that all of the above results smoothly join at the respective boundaries. 

\section{Force--velocity characteristic at strong pinning \label{sec:Fv}}

\begin{figure}[htb]
\includegraphics{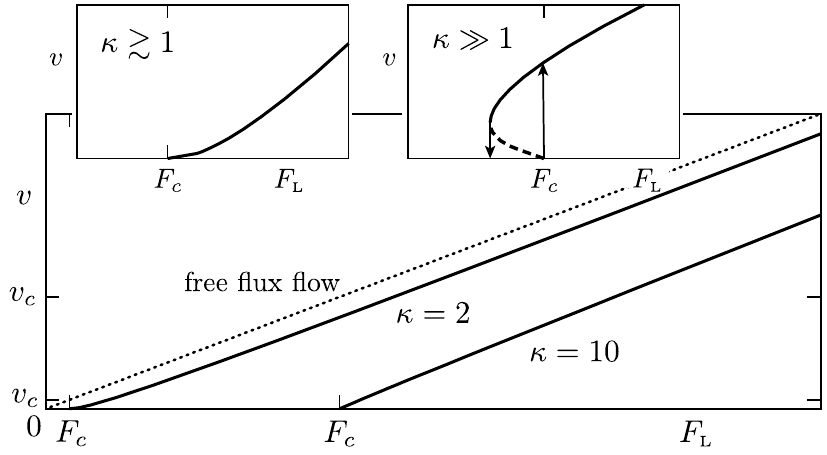}
\caption{\label{fig:dilute_curve} Illustration of the force--velocity curve in
the dilute limit for $n_p a_0 \sigma^2 \kappa = 0.05$.  In the absence of
pinning, the velocity is given by $v=F_{\rm \scriptscriptstyle L}/\eta$
(dotted line). In the presence of pinning, this line is shifted to $F_c\propto
n_p$ (solid line) and closely follows a shifted straight line of equal slope
for velocities $v \ll v_p$.  Corrections to this linear excess-force
characteristic appear at velocities beyond $v_p$, which does not depend on the
small pin density $n_p$, or at small velocities $\propto n_p^2$ (see insets,
the arrows refer to the hysteretic switching). Original figure published in
Ref.\ [\onlinecite{thomann_12}].}
\end{figure}

We are now ready to find the force--velocity or current--voltage
characteristic of the strong pinning superconductor in the dilute pin regime.
The final task is to solve the dynamical equation (\ref{eq:eom}) which we have
already written in the convenient form Eq.\ (\ref{eq:fv}). The analysis of
Section \ref{sec:dynamic} has provided us with the velocity scale $v_p$ for
the average pinning-force density $\langle F_p(v)\rangle$.  In the limit of
small pin densities $n_p$, we find that the dissipative motion of the bulk
vortex system involves the velocity $v_c = F_c/\eta$ which is much smaller
than the typical depinning velocity $v_p$ characteristic of the strong pinning
physics: Interpolating the results \eqref{eq:F_c_vs} and \eqref{eq:F_c_ms} for
the critical force $F_c$, we find that the ratio
\begin{equation}\label{eq:v_ratio}
   \frac{v_c}{v_p} \sim n_p a_0 \sigma^2
   \frac{(\kappa-1)^2}{\kappa} \ll 1
\end{equation}
in the small pin-density limit at fixed $\kappa > 1$.\cite{comment} The
pinning-force density $\langle F_p (v) \rangle$ then remains essentially
unchanged, $\langle F_p (v) \rangle \approx F_c $ for a large region of
velocities including $v_c$ and limited only by $v_p \gg v_c$. Hence, the
characteristic takes the generic form of a shifted (by $F_c$) linear
(flux-flow) curve,
\begin{equation}\label{eq:fv_exc_cur}
   v \approx (F_{\rm \scriptscriptstyle L} - F_c)/\eta,
   \qquad v \ll v_p,
\end{equation}
see Fig.\ \ref{fig:dilute_curve}. The free dissipative flow 
\begin{equation}\label{eq:fv_ff}
   v \approx F_{\rm \scriptscriptstyle L}/\eta,
   \qquad v_p \ll v,
\end{equation} 
is approached only at very high velocities $v \gg v_p \gg v_c$.  

The simple excess-force characteristic is a consequence of the separation of
velocity scales $v_c$ and $v_p$; the latter merge at strong pinning with
increasing density $n_p$ when strong 3D pinning goes over into 1D strong
pinning at\cite{blatter_04} $n_p a_0 \sigma^2 \kappa \sim 1$. Using qualitative
arguments, a similar excess-force characteristic has been found in Ref.\
[\onlinecite{campbell_78}]. 

The roughly constant behavior of the pinning-force density $\langle F_p (v)
\rangle \approx F_c$ over a large velocity region $v <  v_p$ is the analogue
to Coulomb's law of dry friction for the problem of strong vortex pinning.
Hence, although we have invested a large effort in the calculation of the
velocity dependence of the pinning-force density $\langle F_p (v) \rangle$,
the most important statement is that about the largeness of the scale $v_p$ in
comparison to $v_c$. The detailed dependence of $\langle F_p (v) \rangle$ on
$v$ in Eqs.\ \eqref{eq:f_sv_pta} and \eqref{eq:f_sv_ptb} only manifests itself
very close to and far away from $F_c$, e.g., when investigating the approach
to the free flux flow at large drives $F_{\rm\scriptscriptstyle L} \gg F_c$.

Besides the corrections at high velocities $v > v_p$ due to the velocity
dependence of $\langle f_p (v) \rangle$, additional changes show up close
to $F_c$ and at very low velocities due to the square-root dependence $\langle
F_p (v) \rangle -F_c \propto \pm \sqrt{v/v_p}$.  The force balance equation
then can be written in the form
\begin{equation}\label{eq:fsv}
   \frac{F_{\rm\scriptscriptstyle L}}{F_c}-1
   =\frac{v}{v_c}\pm\sqrt{\frac{v}{v_p^\pm}},
\end{equation}
with the small-velocity pinning scales $v_p^\pm$ deriving from Eq.\
(\ref{eq:f_sv_pta}),
\begin{eqnarray}\label{eq:vpm}
   v_p^- &\sim& \frac{\lambda^2}{a_0^2} v_p, 
   \quad \kappa \gg 1 \\ \label{eq:vpp}
   v_p^+ &\sim& \frac{\lambda^2}{a_0^2} (\kappa-1)^4\, v_p,
   \quad \kappa \to 1.
\end{eqnarray}
For strong pinning $\kappa \gg 1$, the negative (non-linear) correction in the
average pinning-force density generates a bistability (and hence hysteretic
jumps\cite{larkin_86}) on the scale $v_\mathrm{nl} = v_c^2/v_p^- \propto
n_p^2$. The unstable branch increases below $F_c$ according to $v \approx
v_p^- (1-F_{\rm \scriptscriptstyle L}/F_c)^2$, turns around reaching a
finite value $v = v_\mathrm{nl}$ at $F_{\rm \scriptscriptstyle L} = F_c$, and
approaches the linear excess characteristic $v \approx v_c (F_{\rm
\scriptscriptstyle L}/F_c-1) > v_\mathrm{nl}$ for $F_{\rm \scriptscriptstyle
L} > F_c$.  On the other hand, approaching the Labusch point $\kappa \to 1$,
the correction changes sign and the velocity increases quadratically,
\begin{equation}\label{eq:vdf2}
   v \sim v_p^+ (F_{\rm \scriptscriptstyle L}/F_c-1)^2 < v_c^2/v_p^+ 
   = v_\mathrm{nl},
\end{equation}
reaches the value $v_\mathrm{nl}$ at $F_{\rm \scriptscriptstyle L}/F_c =
1+2v_c/v_p^+$, and crosses over to the linear regime $v \approx v_c (F_{\rm
\scriptscriptstyle L}/F_c-1) > v_\mathrm{nl}$ at larger drive $F_{\rm
\scriptscriptstyle L} > F_c$. While these features are illustrated in the
insets of Fig.\ \ref{fig:dilute_curve} (showing an expanded view of the
characteristic near onset), we have to caution the reader that these results,
residing in the regime $v_\mathrm{nl} \propto n_p^2$, may get modified due to
collective pinning effects.

\section{Model pins}\label{sec:models}

In our discussion above, we have frequently made use of the Lorentzian-shaped
pinning potential Eq.\ (\ref{eq:e_p_L}) in order to gain insights into the
strong-pinning features of the dynamical vortex-response. This specific
example of a pinning potential is quite appropriate when performing a
numerical analysis but is less convenient for analytical studies. The latter
can be easily attacked for a bare pinning force $f_p(x)$ of polynomial form,
at least in the static limit where we have to solve the self-consistency
equation \eqref{eq:u_s}. It turns out, that for a linear-force profile, the
analytic solution can be pushed further to finite velocities, motivating our
study of pins with a truncated quadratic (or parabolic) pinning potential, see
Fig.\ \ref{fig:linear}.  However, the quadratic-potential pin formally
describes a very strong pin with a Labusch parameter $\kappa \to \infty$,
since the force jumps to zero at the pin's edges at $x = \pm\sigma$.  In order
to study pinning at intermediate values of $\kappa$ and the approach to the
Labusch point $\kappa \to 1$, a more regular potential is required near the
force maximum. The results of such an analysis for a cubic pin (or
quadratic-force model), although analytically accessible in the static limit,
are somewhat cumbersome and we refer the interested reader to Ref.\
[\onlinecite{thomann_thesis}].

\subsection{Parabolic pin, linear-force model\label{subsec:linear_force}}
\begin{figure}[hb!t!]
\includegraphics[width=7cm]{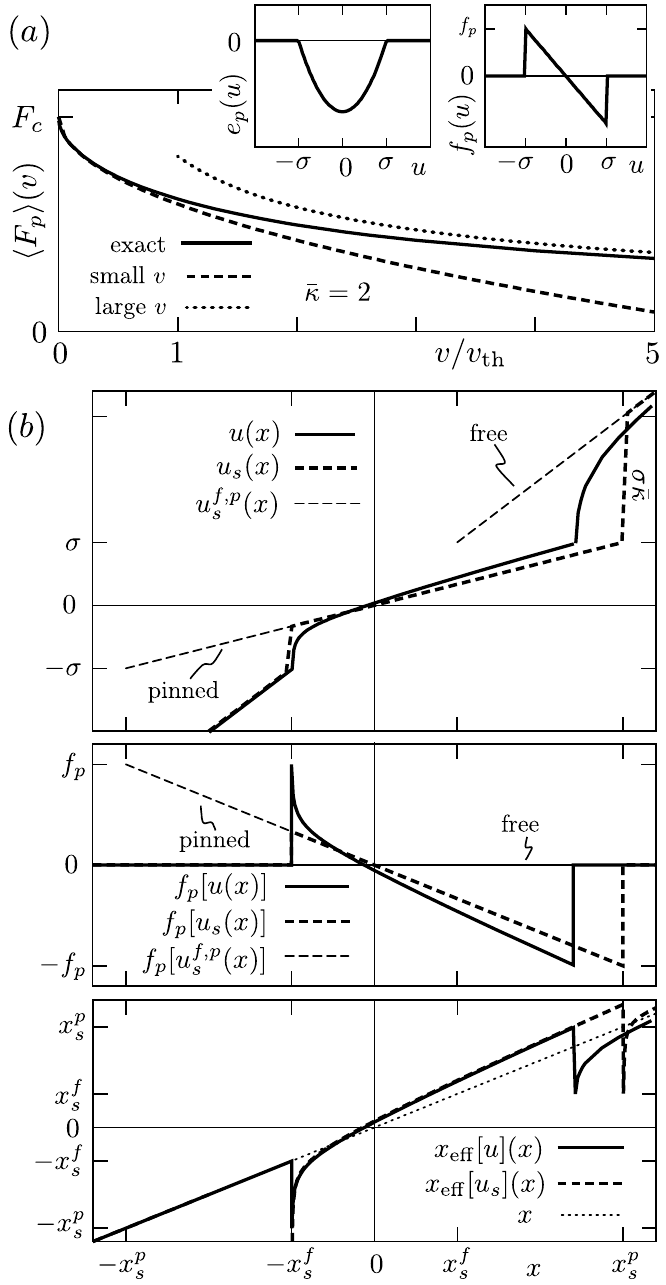}
\caption{\label{fig:linear} a) Pinning-force density versus velocity for a
linear-force pinning model with strength $\bar\kappa = 2$ (the insets show a
sketch of the bare pinning potential $e_p(x)$ and bare force $f_p(x)$). The
exact result (solid) is plotted for comparison with the results from the low-
(dashed) and high velocity (dotted) expansions.  b) from top to bottom:
displacement field $u(x)$, force $f_p[u(x)]$, and effective coordinate
$x_\mathrm{eff}[u](x)$ for a small velocity $v/v_p = 0.05$.  The
exact displacement field $u(x)$ and force $f_p\bigl[u(x)\bigr]$ (solid lines)
are compared to the static quantities $u_s(x)$ and $f_p\bigl[u_s(x)\bigr]$
(dashed) for vortex motion from left to right. Also shown are the free
[$u_s^f(x)$ and $f_p[u_s^f(x)]$, light dashed] and pinned branches [$u_s^p(x)$
and $f_p[u_s^p(x)]$, light dashed], with the multi-valued static solution
within the interval $x_s^f < |x| < x_s^p$.  The exact effective coordinate
$x_\mathrm{eff} [u](x)$ (solid line) is compared to the approximation
$x_\mathrm{eff} [u_s](x)$ (dashed).}
\end{figure}

A simple model for a strong pin is provided by the parabolic potential with the
linear pinning-force restricted to a finite interval $[-\sigma,\sigma]$ 
[see Fig.\ \ref{fig:linear}(a)]
\begin{equation}
   f_p(\myu)=\left\{\begin{array}{l l}-f_p\, \myu/\sigma, & 
   |\myu| <\sigma, \\
   \noalign{\vspace{3pt}}
   0, & \mathrm{otherwise}.
   \end{array}\right.
   \label{eq:linear_force_model}
\end{equation}
Given the jumps in the force at the boundaries $\pm\sigma$, the Labusch
parameter $\kappa = f_p'(\pm\sigma)/\bar{C} \to \infty$ and the pin is strong
for all values of $f_p$. As a consequence, any parabolic potential will
produce vortex pinning, e.g., in numerical simulations \cite{Olson}. On the
other hand, the properties of the pin are determined by the dimensionless
parameter (see Eq.\ (\ref{eq:k}) and the different sign used here)
\begin{equation}
   \bar{\kappa} = -{\partial_x f_p(x)}/{\bar{C}} = f_p/\sigma\bar{C},
   \label{eq:kappa_bar}
\end{equation}
which is of the same scale as the Labusch parameter $\kappa$ for a similar
smooth pin, but should not be confused with the Labusch parameter itself.
Piecewise linear models of this type have been considered before in Refs.\
[\onlinecite{matsushita_79}] using a simplified description of the system's
elastic properties.  Furthermore, Larkin and Ovchinnikov \cite{larkin_86} used
a periodic version of this model in a small-velocity analysis of the
strong-pinning physics for large defects.

In the static situation, we have to solve the self-consistency equation
(\ref{eq:u_s}) for the displacement field $u_s(x)$ and we obtain the result
\begin{eqnarray} \label{eq:lfm_u_s}
   u_s^f(x) &=& x, \quad  |x| > x_s^f = \sigma, \\
   u_s^p(x) &=& \frac{x}{1+\bar{\kappa}}, \quad 
       |x| < x_s^p = \sigma (1+\bar\kappa),
\end{eqnarray}
with the jump points $-x_s^f$ and $x_s^p$ for a right moving vortex separated
by $\sigma (2+\bar\kappa)$. The displacement field $u_s(x)$ then suddenly
changes slope to generate the effective static force $f_p[u_s(x)] = -\bar{C} x
/(1+{\bar\kappa}^{-1})§$.  Note that the jumps at $-x_s^f$ (by $\sigma
\bar\kappa/(1+\bar\kappa)$) and at $x_s^p$ (by  $\sigma \bar\kappa$) do not
disappear for any values of $\bar\kappa$ and hence the pin is always strong.
The critical force Eq.\ \eqref{eq:Fc} assumes the value
\begin{eqnarray}
   F_c&=&n_p f_p \frac{2t_\perp}{a_0^2} \int_{-x_s^f}^{x_s^p} 
   \frac{dx}{\sigma} \,
   \frac{x}{1+ \bar\kappa}
   \nonumber\\
    &=& \frac{2 \bar\kappa \sigma^2}{a_0^2} n_p f_p\, 
   \frac{1+\bar\kappa/2}{1+\bar\kappa},
   \label{eq:linear_critical_force}
\end{eqnarray}
where we have used $t_\perp = x_s^f = \sigma$ in the last equation.  The
factor $2\bar{\kappa}\sigma^2$ should be interpreted as the trapping
area $S_\mathrm{trap}$, see also Refs.\ [\onlinecite{ivlev_91,blatter_04}].

Next, we turn to the low-velocity dynamics (see Fig.\ \ref{fig:linear}) and
determine the coordinate shift $\delta x[u_s](x)$ inside the pinning interval
$[-x_s^f,x_s^p]$, see Eq.\ (\ref{eq:dxus}),
\begin{equation}
   \delta x[u_s](x)=-\frac{2}{\pi} \frac{\sigma\bar\kappa  }{1+\bar\kappa}
   \frac{1+2x/\sigma}{\sqrt{1+x/\sigma}}
   \sqrt{\frac{v}{v_\sigma}}.
   \label{eq:lfm_dxus}
\end{equation}
The divergence at $x=-\sigma$ is cut (at $\sim v t_\mathrm{th}$) by the fast
single-vortex response at short times $t \sim t_\mathrm{th}$ where the 3D
Green's function has to be replaced by the 1D one; the remaining spike is
relevant in transforming the static solution $u_s(x)$ to the dynamic one
$u(x)$. As compared with the result Eq.\ (\ref{eq:dxus}) above, the coordinate
shift for the linear force model does not exhibit any logarithmic corrections
and hence there will be no term $\propto \sqrt{v} \ln v$ in the pinning-force
density $\langle \delta F_p(v)\rangle$. In fact, using the result Eq.\
(\ref{eq:lfm_dxus}) in the calculation of the pinning-force density Eq.\
(\ref{eq:xxs_f_sh_ap}), we obtain
\begin{eqnarray} \nonumber
   \langle \delta F_p(v) \rangle
   &\approx&
   -n_p f_p \frac{2t_\perp}{a_0^2} \biggl[\int_{-x_s^f}^{x_s^p}\!\!
      \frac{dx}{\sigma}\, \frac{\delta x[u_s](x)}{1+\bar\kappa}
   -\delta x[u_s](x_s^p) \biggr] \\ \label{eq:lfm_dFp}
   &=& -\frac{4}{3\pi}\frac{\bar\kappa \sigma t_\perp}{a_0^2}
   n_p f_p \frac{5+5 \bar\kappa+2\bar\kappa^2}
                             {(1+\bar\kappa)^{2}}
   \sqrt{\frac{v}{v_p}},
\end{eqnarray}
with $v_p = v_\sigma (2+\bar\kappa)$ deriving from the effective pin size
$\sigma(2+\bar\kappa)$. As discussed above, the pinning force is reduced
$\propto \sqrt{v/v_p}$ with respect to the static critical force in the strong
pinning situation discussed here. A result similar to Eq.\ (\ref{eq:lfm_dFp})
was obtained by Larkin and Ovchinnikov \cite{larkin_86} in a periodic linear
model for large defects.

\begin{figure}[hb!t!]
\includegraphics{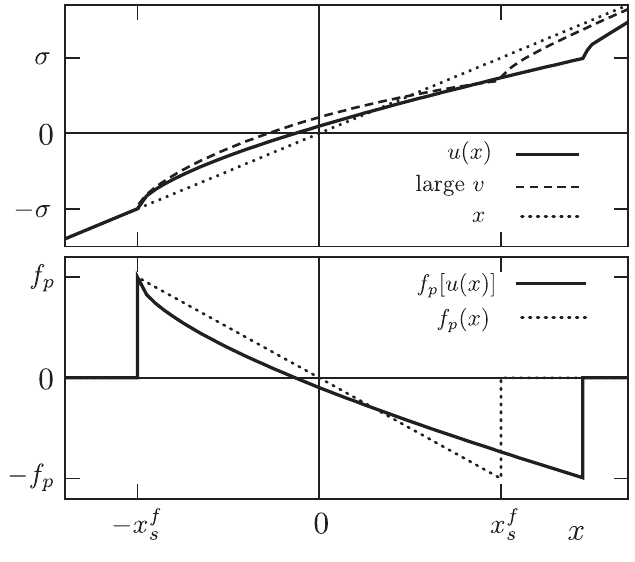}
\caption{\label{fig:linear_fast} Displacement field $u(x)$ and corresponding
force $f_p[u(x)]$ at a high velocity $v/v_p = 2$ and for
a pinning parameter $\bar{\kappa}=2$. The exact results (solid lines) are
compared with the high-velocity perturbative (dashed) and free (light dotted)
results.}
\end{figure}

Going to the limit of high velocities $v \gg \bar{\kappa} v_p$, we
first determine the lowest-order correction to the displacement field, see
Eq.\ (\ref{eq:duhv}),
\begin{eqnarray} \label{eq:lfm_du}
   u_p(x)&=&-\frac{2\, \bar{\kappa}}{\pi}
   \frac{\partial}{\partial x}
   \int_{-\sigma}^x dx' x' \sqrt{\frac{x-x'}{v t_\mathrm{th}}}
   \\\nonumber
   &=&-\frac{2\, \bar{\kappa} \sigma}{3\pi} \sqrt{\frac{v_\sigma}{v}}
   \Bigl(\frac{2x}{\sigma}-1 \Bigr) \sqrt{1+\frac{x}{\sigma} }
\end{eqnarray}
for $-\sigma<x<\sigma$, see Fig.\ \ref{fig:linear_fast}. Using the
perturbative expression  Eq.\ \eqref{eq:F_pt} for the pinning force (note that
$f_p'(x)$ includes a $\delta$-function), we arrive at the result (cf.\ Eq.\
\eqref{eq:fpt_lv})
\begin{eqnarray}
   \langle F_p (v) \rangle &\approx &
   \frac{12\sqrt{2} }{5\pi}\, \frac{\sigma t_\perp}{a_0^2} n_p f_p 
   \sqrt{\frac{\bar{\kappa}^2 v_\sigma}{v}}.
   \label{eq:lfm_F_p_hv}
\end{eqnarray}

The linear-force model also allows for an {\it exact} solution of the
dynamical situation; within the interval $u \in [-\sigma,\sigma]$ the
self-consistency equations \eqref{eq:u_s} and (\ref{eq:d_x}) assume the form
\begin{eqnarray}
   \myu(x)&=&\frac{x-\delta x[\myu](x)}{1+\bar\kappa},
   \label{eq:static_dynamic_equation_linear}\\
   \delta x[\myu](x)&=&-\frac{f_p}{\sigma}\frac{\partial}{\partial x}
   \int_{-\sigma}^x dx' G^\uparrow(x-x')\myu(x').
   \label{eq:lfm_u_dx}
\end{eqnarray}
This problem can be solved exactly via a Laplace transform and we obtain the
solution\cite{LT}
\begin{eqnarray}
   \label{eq:LT}
   u(z)&=&\frac{1}{z^2 (1 + \bar\kappa g(z))},\\
   g(z) &=& e^{zvt_\mathrm{th}} \mathrm{erfc}(\sqrt{zvt_\mathrm{th}}),
\end{eqnarray}
with $\mathrm{erfc}(x)$ the complementary error function.  Unfortunately, the
final solution involves an inverse Laplace transform which cannot be done in
closed form,
\begin{eqnarray}
   \label{eq:LT_u}
   u(x)&=& \frac{x-\bar\kappa \sigma}{1+\bar\kappa} +
   \sigma \bar\kappa \int_0^\infty \frac{ds}{\pi} 
   \Bigl(\frac{v/v_\mathrm{th}}{s^2} +\frac{1}{s}\Bigr) \\
   \nonumber && \quad \times
   \frac{(1-e^{-s(1+x/\sigma)v_\sigma/v})e^{-s} \mathrm{erfi}(\sqrt{s})}
   {(1+ \bar\kappa e^{-s})^2 +\bar\kappa^2 e^{-2s} \mathrm{erfi}^2(\sqrt{s})},
\end{eqnarray}
where $\mathrm{erfi}(x)$ denotes the imaginary error function.  The result of
the numerical evalution of \eqref{eq:LT_u} is shown in  Fig.\ \ref{fig:linear},
together with the results of the fast- and the slow-velocity analysis.

\begin{figure}[ht!b!]
\includegraphics{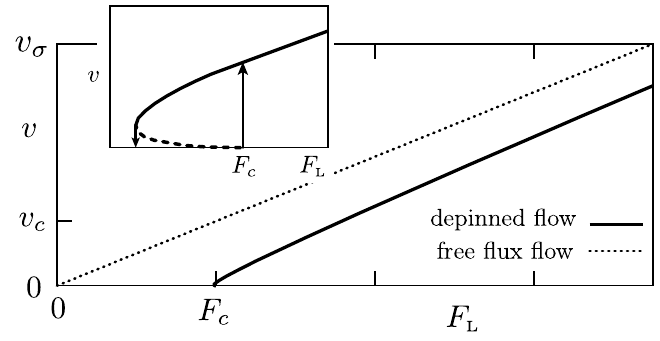}
\caption{\label{fig:linear_resulting_force} The exact force-density--velocity
(current--voltage) relation for parabolic pins with strength $\bar\kappa=f_p
/\sigma\bar{C}=2$. The pinning velocity scale $v_p = 2 v_\sigma$ is much
larger than the viscous velocity scale $v_c=F_c/\eta$. Note the bistability
for values of the Lorentz force density $F_{\rm\scriptscriptstyle L}$ close to
the critical force-density $F_c$ (inset), leading to the appearance of jumps
\cite{larkin_86}.  This bistable regime extends over a region of size $\propto
n_p^2$ both in force and in velocity. For large values of
$F_{\rm\scriptscriptstyle L}$, free flux flow (dashed) is approached.}
\end{figure}

In the final step, we make use of the mean pinning-force density $\langle F_p
(v) \rangle$ in the solution of the force-balance equation \eqref{eq:eom} and
find the force--velocity characteristic, see Fig.\
\ref{fig:linear_resulting_force}. For small velocities $v\ll v_p$, the force
balance equation assumes the form
\begin{equation}
   \eta v+ F_c (1-\beta\sqrt{v/v_p})
   =F_{\rm\scriptscriptstyle L},
   \label{eq:linear_force_balance}
\end{equation}
with
\begin{eqnarray}\label{eq:beta}
   \beta &=& \frac{4}{3\pi}
   \frac{5+5\bar\kappa+2\bar\kappa^2}{(2+\bar\kappa)(1+\bar\kappa)}
   \stackrel{\bar\kappa \to \infty}{\longrightarrow}
   \frac{8}{3\pi}
\end{eqnarray}
determined from Eq.\ \eqref{eq:lfm_dFp}. The pinning-force density, decreasing
with velocity via a square-root law, outperforms the linear behavior of the
viscous force density $\eta v$ at small $v$.  As a consequence, the
force--velocity relation is bistable with (we define the small parameter $\nu
= \beta^2 v_c/4v_p \sim n_p \bar\kappa\sigma^2a_0 \ll 1$)
\begin{equation}
   \frac{v}{v_c}
   \approx \nu 
   \left(1\pm\sqrt{1+\frac{1}{\nu} \frac{F_{\rm\scriptscriptstyle L}-F_c}
   {F_c}}\right)^2.
   \label{eq:lfm_sv}
\end{equation}
Eq.\ (\ref{eq:lfm_sv}) describes an unstable characteristic $v/v_c\approx
(1/4\nu)(1-F_{\rm\scriptscriptstyle L}/F_c)^2$ at small $v$ and a jump $\delta
v/v_c \approx 4\nu$ at $F_{\rm\scriptscriptstyle L} = F_c$, see Fig.\
\ref{fig:linear_resulting_force}, as also noted by Larkin and Ovchinnikov in
their pinning analysis of large defects \cite{larkin_86}.  The bistability
regime extends over a region of size $\propto n_p^2$ both along the force- and
along the velocity axes; specifically, bistability appears within the force
interval $1-\nu < F_{\rm \scriptscriptstyle L}/F_c < 1$ and for velocities
$v/v_c < 4\nu$.  The electric field $E_c$ corresponding to the jump velocity
$4 \nu v_c$ can be expressed through the critical current density $j_c$ and
the flux-flow resisitivity $\rho_\mathrm{ff}$, $E_c = 4 \nu \rho_\mathrm{ff}
j_c \sim (\kappa n_p \sigma^2 a_0) \rho_\mathrm{ff} j_c$, with $\kappa n_p \sigma^2
a_0 \ll 1$ assuring the dilute pinning limit.

Above the critical drive $F_{\rm\scriptscriptstyle L}-F_c \ll F_c$, we find
the linear shape $v/v_c \approx F_{\rm\scriptscriptstyle L}/F_c-1$.  Hence,
the force--velocity relation is dominated by the static pinning-force density
$F_c\propto n_p$ and the viscous force density $\eta v$.  For large
velocities, we use $\langle F_p (v) \rangle$ from the expansion Eq.\
\eqref{eq:lfm_F_p_hv} and find the correction around flux flow
\begin{equation}
	v\sim \frac{F_{\rm\scriptscriptstyle L}}{\eta}
	-\frac{4 \sqrt{2}}{3\pi}
	\frac{\bar\kappa \sigma t_\perp}{a_0^2} \frac{n_p f_p}{\eta}
	\sqrt{\frac{v_\sigma}{v_c}\frac{F_c}{F_{\rm\scriptscriptstyle L}}}.
	\label{eq:lfm_hv}
\end{equation}

The linear-force model always resides in the strong pinning limit with $\kappa
= \infty$. The most relevant point in the pinning process, the point $x_s^p =
\sigma (\bar\kappa+1)$ where the vortex jumps out of the pin, then coincides
with the point of maximal (negative) force with $u_s(x_s^p) = \sigma$ at the
pin boundary. In order to analytically study the dynamical pinning force at
smaller values of $\kappa$ including the approach to the Labusch point, one
has to choose a smooth shape for the pinning potential. Demanding that the
static limit is still analytically solvable, one may choose a cubic potential
with the quadratic force profile
\begin{equation}
   f_p(x)=f_p \left\{
   \begin{array}{l l}
   \displaystyle
   -(x/\sigma)^2-x/\sigma,
   & -\sigma< x <0,\\\\
   \displaystyle
   (x/\sigma)^2-x/\sigma,
   & 0< x <\sigma,
   \end{array}
   \right.
\label{eq:parabolic_force}
\end{equation}
and zero otherwise. This model pin then comes with a tunable Labusch
parameter
\begin{equation}
   \kappa=\max_{x} \frac{\partial_x f_p(x)}{\bar{C}}
   =\frac{\partial_x f_p(\pm\sigma)}{\bar{C}}
   =\frac{f_p}{\sigma\bar{C}}
\end{equation}
and can still be solved analytically in the static limit, that provides the
basis for an approximate solution of its dynamical behavior.  Since the
results are rather cumbersome, we refer the reader to the original solution in
Ref.\ [\onlinecite{thomann_thesis}].

\section{Conclusion}\label{sec:conclusion}
We have analyzed the dynamics of the vortex lattice in the presence of dilute
strong pins characterized by a Labusch parameter $\kappa>1$ and have
determined the strong-pinning force--velocity (or current--voltage)
characteristic in the single-pin single-vortex limit. The basic task is the
solution of a nonlinear integral equation for the displacement field $u(x=vt)$
describing the vortex tip position when traversing the pin while the vortex
ends at $z \sim \pm \infty$ move with constant velocity $v$. The average
$\langle f_p(v)\rangle$ over the individual pinning forces $f_p[u(x)]$ and a
proper determination of the velocity dependence of the transverse trapping
length $t_\perp(v)$ provide the velocity-dependent mean pinning-force density
$\langle F_p (v) \rangle$.  In a last step, we find the mean velocity
$v(F_{\rm\scriptscriptstyle L})$ as a function of the driving Lorentz-force
density $F_{\rm\scriptscriptstyle L}$---the force--velocity or
current--voltage characteristic---by solving the force balance equation $\eta
v = F_{\rm\scriptscriptstyle L} - \langle F_p (v) \rangle$.

The self-consistent dynamical integral equation for the displacement field
$u(x)$ can be solved numerically by simple forward integration due to
causality. Such a numerical solution has been carried out for the
Lorentzian-shaped pin in order to determine the dynamical effective pinning
force $f_p[u(x)]$ and the dependence of the pinning force $\langle f_p (v)
\rangle$ on the velocity $v$. In the static limit, the integral can be
separated and the problem simplifies to an algebraic one. The appearance of a
multi-valued static solution characterizes a strong pin and allows for a
finite static critical force $f_c = \langle f_p (v=0) \rangle$; the latter is
a consequence of the asymmetric occupation of free and pinned branches as the
moving vortex jumps into and out of the pin. At finite velocities, the jumps
in the static solution $u_s(x)$ give way to a unique smooth and asymmetric
solution $u(x)$ and the force $f_p[u(x)]$ derives from a direct integration
without invoking an asymmetric occupation.

The velocity dependence of the dynamical pinning force $\langle f_p (v)
\rangle$ is governed by the elastic properties of the vortex system as
expressed through the Green's function. Its velocity scale is given by $v_p =
\kappa v_\sigma = \kappa \sigma/t_\mathrm{th}$, where $\kappa$ and $\sigma$
encode properties of the pins and $t_\mathrm{th}$ is the timescale for the
dissipative relaxation of an elastic deformation.  At small velocities, a
perturbative treatment away from the static solution provides a decrease
$\langle \delta f_p(v)\rangle \propto - \sqrt{v/v_p}$ of the pinning-force
density at large values of $\kappa \gg 1$ and an increase $\langle \delta
f_p(v)\rangle \propto \sqrt{v/v_p}$ for small values $\kappa \gtrsim 1$.
Indeed, a correction of positive sign $\langle \delta f_p(v)\rangle \propto
\sqrt{v/v_p} \log(v_p/v)$ shows up in a small region as $v \to 0$ for any
value of $\kappa$, however, this region is exponentially small at large
$\kappa$.  Increasing the velocity beyond $(a_0/\lambda)^2 v_p$, the 3D bulk
response gives way to a region of 4D dispersive behavior and $\langle \delta
f_p(v)\rangle \propto v/v_p$ changes linearly with velocity. At high
velocities $v > v_p$, the rapid motion is dominated by the 1D single-vortex
response; furthermore, a self-consistent analysis of the problem shows that
the longitudinal trapping or pinning length decreases from $\sim \kappa
\sigma$ to $\sim \sigma$, producing a rapid drop in the pinning force $\langle
\delta f_p(v)\rangle \rangle \propto v_p/v$. The $v_p/v$ decay of $\langle f_p
(v)\rangle$ extends over a large velocity regime $v_p < v < \kappa v_p$ and
has not been known before.  Finally, perturbation theory away from free
flux flow can be performed at high velocities $v \gg \kappa v_p$ where pinning
is always effectively weak and which provides a generic correction $\langle
\delta f_p (v)\rangle \propto (\kappa v_p/v)^{1/2}$.

The pinning-force density $\langle F_p(v)\rangle \approx n_p
(2t_\perp(v)/a_0)$ $\langle f_p (v)\rangle$ follows from simple averaging via
direct summation over independent pins, once the dynamical transverse pinning
length $t_\perp(v)$ has been determined. The latter can be found by studying
the trapping process for a vortex approaching the defect with a finite impact
parameter and at finite velocity; it turns out that the trapping length
decreases from its static value $t_\perp \sim \sigma \kappa^{1/(n+2)}$ to
$t_\perp \sim \sigma$ as the velocity increases to $v_p$.

Our study provides access to several types of results, i) exact ones deriving
from numerical integration for specific defect potentials $e_p(r)$ and ii)
perturbative analytic results in the static- and high-velocity limits. Quite
remarkable are, iii) the universal results obtained in the large-$\kappa$
limit, with a linear force $f_p[u_s(x)] \approx - \bar{C}x$ appearing in the
static limit and a square-root force $f_p[u(x)] \approx - (2/\pi) \bar{C}
\sqrt{v t_\mathrm{th} x}$ in the intermediate velocity regime $v_p < v <
\kappa v_p$. Finally, iv) dimensional estimates provide us with simple
qualitative results for the evolution of $\langle F_p(v)\rangle$ with changing
velocity $v$ and for all values of $\kappa$.  v) Numerical and analytic
calculations for a parabolic model-potential give additional insights into the
pinning dynamics. Such parabolic pins have often been used in numerical
simulations of vortex dynamics in disorder landscapes \cite{Olson} and always
reside in the strong-pinning limit due to their sharp boundary.

The analysis of the velocity-dependent pinning-force density $\langle F_p (v)
\rangle$ provides us with the velocity scale $v_p$ governing the relaxation of
the vortex motion across the pinning centers.  In the limit of a dilute
density $n_p$ of pins (where pinning centers act individually) this velocity
is much larger than the velocity scale $v_c = F_c /\eta$ describing the
overall motion of the vortex system after depinning. It is this separation of
velocities that produces a simple force--velocity or current--voltage
characteristic in the strong pinning situation with a dilute density of pins:
after depinning at $F_c$ (or $j_c$), the characteristic evolves first in
parallel to free flux-flow (excess-force characteristic $v \sim
(F_{\rm\scriptscriptstyle L}-F_c)/\eta$) and approaches the free flow behavior
$v \sim F_{\rm\scriptscriptstyle L}/\eta$ only at very high velocities $v >
v_p \gg v_c$. With our derivation of the excess-force characteristic for the
strongly-pinned vortex system in the dilute-pin limit, we have derived the
analogue of Coulomb's law of dry friction for strong vortex pinning---it would
be interesting to see if the ideas leading to this result could be applied to
other problems of dry friction.  Additional nonlinear corrections
(corresponding to corrections of Coulomb's law), a jump at very strong pinning
with $\kappa \gg 1$ and a smooth onset for $\kappa \gtrsim 1$ appear in a
narrow region of size $\propto n_p^2$ near depinning; however, a word of
caution is in place as results in this regime may get modified by collective
pinning effects.

Our theoretical results compare well to a number of experimentally measured
current--voltage characteristics
\cite{strnad_kim_64_65,huebener_79,berghuis_93,zeldov_02,pace_04}.  The linear
excess-current characteristic reported in early experiments was discussed by
Campbell and Evetts\cite{campbell_72} and by Campbell\cite{campbell_78},
however, no `microscopic' derivation of this basic result has been provided so
far. Unfortunately, one has to admit that even today, no systematic studies of
experimental current--voltage characteristics are available: Given a specific
material, the defect structure is usually non-trivial and may include a
variety of pin types.  Furthermore, the parameters characterizing the defects
are difficult to find.  Experiments with superconductors where defects could
be designed, tuned, and properly characterized would provide a great help and
motivation in further developing the theory of pinning and allow a better
comparison between theory and experiment. Numerical studies based on the
time-dependent Ginzburg-Landau theory and aiming at optimizing the pinning
landscape have been performed very recently \cite{Sadovskyy} and it will be
interesting to compare our results with this type of numerical effort.

Further work is required on the theoretical side: With our analysis,
we have provided an important step in the understanding of the static and
dynamical strong pinning behavior and its crossover to weak pinning.  However,
our study is limited to the single-pin single-vortex situation and one has to
include effects of other vortices and correlations between pins in order to
arrive at a complete picture. Correlations between pins will generate
higher-order corrections in $n_p$ both in the static and the low-velocity
dynamic behavior and their inclusion is a crucial element in the full
understanding of the weak to strong pinning crossover.\\

\begin{acknowledgments}
In memoriam of Anatoli Larkin who has initiated this study. We thank Roland
Willa and Martin Buchacek for numerous discussions and acknowledge financial
support of the Fonds National Suisse through the NCCR MaNEP.
\end{acknowledgments}




\begin{thebibliography}{99}
\bibitem{indices} Note that we distinguish defect and vortex positions only by
roman and greek indices, implying that these are not to be replaced by equal
numbers.
\bibitem{com_nu} Note that a precise determination of $\nu$ requires a
numerical analysis, see the discussion in Ref.\ [\onlinecite{willa_16}].
\bibitem{comment} Note that the limit $\kappa \to 1$ at fixed $n_p$ would take
us out of the single-pin regime: while $F_c^{\rm \scriptscriptstyle SP}(\kappa
\to 1) \to 0$, $F_c^{\rm \scriptscriptstyle WCP}$ stays finite. The condition
to remain in the strong pinning regime then reads $F_c^{\rm \scriptscriptstyle
SP} > F_c^{\rm \scriptscriptstyle WCP}$, which requires that
$(\kappa-1)^2/\kappa > (a_0/\lambda_0)^2 (f_p/\varepsilon_0)^3 (n_p
a_0^4/\xi)$.
\bibitem{LT} We define the Laplace transform as $f(z) = \int_0^\infty
dx \,\exp(-zx) f(x)$ and its inverse via $f(x) = \int_{\gamma -i\infty}^
{\gamma+i\infty} (dz /2\pi i) \exp(zx) f(z)$.
\end{thebibliography}
\end{document}